\documentclass[
]{ceurart}

\usepackage{xcolor}

\newcommand\code[1]{{\small\texttt{#1}}}

\usepackage{graphicx}
\usepackage[ruled,vlined]{algorithm2e}

\SetCommentSty{mycommfont}

\usepackage{algorithmic}
\usepackage{xcolor, colortbl}
\usepackage{color}
\usepackage{amsmath,amssymb,amsfonts}
\usepackage{subcaption}
\newtheorem{heuristic}{Heuristic}
\usepackage{booktabs}
\usepackage{multirow}
\usepackage{array}
\begin{document}

\copyrightyear{2022}
\copyrightclause{Copyright for this paper by its authors.
  Use permitted under Creative Commons License Attribution 4.0
  International (CC BY 4.0).}

\conference{CAMLIS'22: Conference on Applied Machine Learning in Information Security (CAMLIS),
  October 20--21, 2022, Arlington, VA}

\title{Inroads into Autonomous Network Defence using Explained Reinforcement Learning}

\author[1]{Myles Foley}[%
email=m.foley20@imperial.ac.uk
]
\author[1]{Mia Wang}[%
email=yixuan.wang18@imperial.ac.uk
]
\address[1]{Imperial College London}

\author[2]{Zoe M}[%
email=zm@turing.ac.uk,
]

\author[2]{Chris Hicks}[%
email=c.hicks@turing.ac.uk,
]
\author[2]{Vasilios Mavroudis}[%
email=c.hicks@turing.ac.uk,
]
\address[2]{The Alan Turing Institute }

\newif\iflong\longfalse
\newcommand\longv[1]{\iflong {\color{blue} #1}\else\fi}

\begin{abstract}
Computer network defence is a complicated task that has necessitated a high degree of human involvement. However, with recent advancements in machine learning, fully autonomous network defence is becoming increasingly plausible. 
This paper introduces an end-to-end methodology for studying attack strategies, designing defence agents and explaining their operation. First, using state diagrams, we visualise adversarial behaviour to gain insight about potential points of intervention and inform the design of our defensive models. We opt to use a set of deep reinforcement learning agents trained on different parts of the task and organised in a shallow hierarchy. Our evaluation shows that the resulting design achieves a substantial performance improvement compared to prior work. Finally, to better investigate the decision-making process of our agents, we complete our analysis with a feature ablation and importance study.
\end{abstract}


\maketitle

\section{Introduction}\label{sec:intro}
Computer network security is characterised by an asymmetry as 
the defender needs to ensure constant protection of the network's components, 
while the adversary can opportunistically single-out weak entry points. 
Such asymmetries have been identified and addressed in many other areas of 
cyber security. For example, cryptographic protocols (e.g., TLS) thwart 
denial of service attacks by ensuring that the prover commits enough
computation cycles before the verifier does so. 
In network defence, however, the problem remains open as 
the task is complex~\cite{speicher_towards_2019} and involves
a wide array of both attack vectors and mitigation tools. Thus, 
network defence is currently handled primary by human experts which entails
high operational costs. 


RL, and particularly deep RL (DRL), excels in interactive tasks that cannot easily be solved using analytical solutions. Human and even super-human levels of performance have been achieved in a range of complex tasks including classic board games such as chess and Go \cite{mnih2016asynchronous, Mnih2015}, video games ranging from classic Atari \cite{mnih_playing_2013,schulman_proximal_2017} to multi-player real-time strategy games \cite{openai_et_al_dota_2019}, autonomous driving \cite{sallab_deep_2017}, and robotics \cite{kober_reinforcement_2013}. Recently, DRL has also been successfully applied to autonomous network defence \cite{foley_autonomous_2022}, a highly interactive task where the defender proactively monitors the state of the network, identifies abnormalities, and acts to remediate them. Commonly, this takes the form of a shallow hierarchy of \emph{specialised subagents} coordinated by a \emph{controller}, any combination of these being autonomous.


To date however, there has been limited consideration for the explainability of these models. Explainable AI has, in domains such as natural language processing and computer vision \cite{williford_explainable_2020}, proven useful not only for end users but also experts and developers of AI systems. DRL models are particularly challenging to explain because the neural networks which represent their agent policies are not readily understandable by humans. Nonetheless, the ability to explain and understand the actions of an autonomous defensive agent is critical. This work investigates, and answers in the affirmative, whether explainable RL (XRL) models and environments can improve autonomous defensive capabilities and aid in their development. 

\subsubsection{Contributions}
Our main contributions are:

\begin{itemize}
    \item We develop methodologies for visualising (i.e., explaining) attacker functionality in the CybORG cyber environment. Our methodology highlights previously undocumented differences in the adversary models and motivates two new controller architectures with improved classification accuracy.
    
    \item We present the full details of our new controller and specialised subagent models. We then evaluate them against two classes of adversary in the CybORG environment realising substantial performance improvements. 
    
    \item We perform a feature ablation and importance study to understand the most influential elements in the observation space and explain our model outputs. 

    
\end{itemize}

\section{RL Background}

In this section we discuss the key RL techniques that are relevant for the rest of the paper. 

\subsection{Deep RL Algorithms}

\subsubsection{PPO}
Proximal Policy Optimisation (PPO) is an efficient policy gradient method  ~\cite{schulman_proximal_2017} for DRL. It has been shown to outperform other popular algorithms such as A3C~\cite{mnih2016asynchronous}, achieving super-human performance in a variety of complex tasks including 49 separate ATARI arcade games~\cite{schulman_proximal_2017}. Despite its effectiveness in very complex environments~\cite{yu_surprising_2022}, it has seen only limited use in security settings~\cite{nguyen_deep_2021, wu_adversarial_2021}, 

PPO uses a policy $\pi_\theta$ ($\theta \in \mathbb{R}$) with an objective function that is  defined by the total reward ${J(\theta) = \mathbb{E}_{\pi_0}[\sum^{\infty}_{t=0}\gamma^t r_t]}$. By formulating the objective function in this way actor-critic architectures can be used: the actor selects an action which is evaluated by the critic. The policy gradient is then computed:  
\begin{equation}
    \bigtriangledown_\theta J(\theta) = \mathbb{E}_{\pi_0}[\bigtriangledown_\theta \log \pi_\theta (s, a)A_{\pi_\theta}(s)]
\end{equation}
$A_{\pi_\theta}(s)$ is the advantage of taking action $a$ instead of the \emph{average} action as computed by the policy $\pi_\theta$ ($Q_{\pi_\theta}(s,a)x-V_{\pi_\theta}(s)$)~\cite{sutton_reinforcement_2018}.

During gradient descent, PPO introduces a clipping function to both prevent reaching local optima during large updates and avoid smaller updates that significantly increase the length of training.

\subsection{Curious Exploration}

Curiosity is a technique that enables agents to explore their environment based on an intrinsic reward signal not provided by the environment~\cite{ICM_curiosity_17}. Such a signal is particularly useful in the absence of a continual extrinsic reward (e.g., the running score found in some games).
Pathak et al. \cite{ICM_curiosity_17} introduce the Intrinsic Curiosity Module (ICM), a semi-supervised technique in which agents choose actions based on the uncertainty in the outcome of each action, intrinsically motivating the exploration of unknown states. ICM also ensures that agents are only incentivised to reach states that are impacted by their actions, avoiding those which are inherently unpredictable. 


\longv{    
\subsection{Multi-Armed Bandits}
\label{ssec:mab}
        The Multi-Armed Bandit (MAB) is an RL problem so called after `one-armed bandits' or slot machines. It is a special case of the markov decision process where all actions return to the same state, thus there is only one state. In such problems agents take a single action and observe the reward, the goal is then to maximise this reward. In this way it can be seen to have an episode length of \(1\), and a maximum number of episodes \(M\). Thus the agent must explore the actions to find the optimal strategy to maximise this reward over \(M\). \cite{sutton_reinforcement_2018}
}

\subsection{Explainable RL}
Explainable RL (XRL), a fledgling sub-field of explainable AI, is the study of tools and methods which enhance human understanding of the actions taken by autonomous agents. A recent and thorough review of XRL is provided by Heuillet at al.~\cite{HEUILLET2021106685} and separately by Puiutta and Veith~\cite{puiutta20XRL}. XRL methods are commonly divided between those which are intrinsic, sometimes called transparent, and those which are post-hoc. Intrinsic XRL models are inherently interpretable and offer explainability at the time of training. In contrast, post-hoc explainability occurs after training; often by creating a second, simpler model to provide explanations. In DRL, learned policies are represented by neural networks making them difficult to interpret. Post-hoc explainability allows the performance advantages of DRL~\cite{Mnih2015} to be retained whilst facilitating human understanding of autonomous decision making. Explainability is not limited to users and experts affected by the decisions of models but, as in this work, is a valuable researcher's aid in developing more efficient and higher-performance models. 

\section{Network Simulation Environment}\label{sec:netsim}
\begin{figure}
    \centering
    \includegraphics[width=0.7\textwidth]{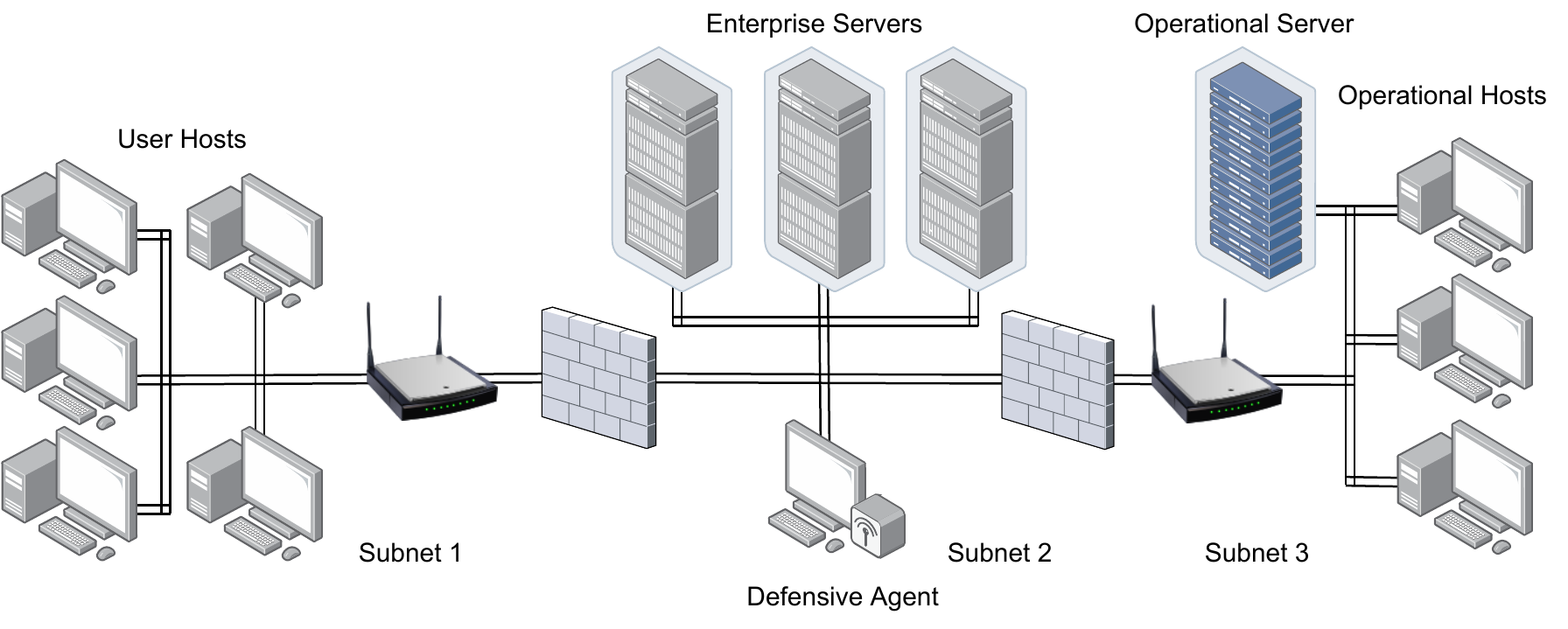}
    \caption{The CybORG environment showing the three subnets and their corresponding hosts and firewalls.}
    \label{fig:cage_2}
\end{figure}

We use the CybORG environment~\cite{cyborg_acd_2021} which simulates the computer network of a manufacturing plant, as shown in Figure~\ref{fig:cage_2}. The network consists of five \textit{user hosts} (Subnet 1), three \textit{enterprise servers} (Subnet 2\footnote{Subnet 2 also includes the defender's machine.}), three \textit{operational hosts} and the \textit{operational server} (Subnet 3). Each host exposes a number of network services that other hosts can connect to, and which may have exploitable vulnerabilities. However, due to the network's firewalls hosts in Subnet 1 cannot directly connect to machines in Subnet 3, and the operational server is accessible only through the operational hosts. The liveness of the operational server has a direct impact on the manufacturing and is considered critical. CybORG assumes two players, a defender and an adversary, who interact with the turn based environment using the actions available to them. 


A common drawback of simulated environments in RL is the \textit{reality gap} which causes agents not to generalise sufficiently when moved from the simulation (i.e., training) to reality (i.e., evaluation). This is due to the simulation not adequately matching reality (e.g., in robotics). To address this, CybORG provides a network \textit{emulator} that runs on Amazon Web Services (AWS). The combination of simulation and emulation ensure that the reality gap is minimised, with the actions available and their effect on the environment consistent across both~\cite{cyborg_acd_2021}. 

The CybORG environment is host to the `Cyber Autonomy Gym for Experimentation' (CAGE) challenge~\cite{cage_challenge_1, cage_challenge_2,cage_challenge_announcement}. CAGE is an international Kaggle-style competition, providing an increasingly challenging benchmark for the evaluation of autonomous defensive agents. The competition is currently in its second iteration (CAGE II).


\subsection{Action Space}
Attackers and defenders have unique action spaces. Defenders perform actions at the host level: 1) Analysing the processes running, 2) Terminating malicious processes, 3) Restoring the host to a previous (benign) state, and 4) Deploying \textit{honeypot}\footnote{Honeypot refers to a decoy system or service that lures attackers by appearing to suffer from known security vulnerabilities. Honeypots are used to detect malicious actors and study their behaviour.} services.
Adversaries can: 1) Scan a subnet for hosts, 2) Scan the ports of a host, 3) Exploit a service on a port, 4) Escalate their access, and 5) Disrupt the services on the operational server. Both players have a `sleep' action to perform no action on the network. Based on the selected actions, the environment updates its \emph{state} and updates the agents' scores. It should be noted that even valid actions may not succeed, as the CybORG simulator introduces randomness to mimic the behaviour of the emulator (e.g. a valid node restoration may occasionally fail).

\subsection{Observation Space}
The defender's observation space is a vector of 52~bits i.e., 4 bits for every network host. The first two bits represent whether the host state is unknown (none), scanned or exploited (set when a decoy is triggered); the last 2 bits specify the access the attacker has on the host machine (i.e., none, user and administrator). 

As in a real network defence situation, neither the defender nor the adversary is omniscient. Neither agent knows the state of the network or the other's position with absolute certainty. In addition, the outcomes of actions are stochastic. For example, from the defender's perspective, when an exploit fails it is not possible to precisely determine which exploit was attempted. This can be crucial information in the instance that an adversary favours a specific exploit strategy. A better informed defender could strategically place decoys on the targeted service to frustrate and evade further attempts more effectively. 


\subsection{Reward Function}
Most games include a scoring function that quantifies the performance of the player. Similarly, CybORG uses a reward function that rewards the adversary and penalises the defender for every compromised or impacted network host. The reward function is as follows: on each turn, for every host on which the adversary has admin access, the defender receives a reward of -0.1 and for every server the reward is -1. There is a -10 reward for disruption on the operational server and a -1 reward when any device is `restored'. In the context of RL, the negative reward for the defensive agent incentivises the agent to take actions that minimise the effect of the adversary.





\subsection{Adversaries}\label{ssec:adversaries}
The environment includes two adversaries: the \code{BLineAgent} that has prior knowledge (i.e., full knowledge of the network's structure but not its current state), and the \code{MeanderAgent} which does not have any prior information. Both agents share the same objective, to reach the operational server and, after escalating their privileges, disrupt its services (i.e., impact its liveness). Due to prior knowledge, the \code{BLineAgent} follows an optimal exploitation trajectory to the operational server. In contrast, the \code{MeanderAgent} breadth-wise scans the network for vulnerable hosts and gradually traverses the subnets. To prevent trivial defence strategies, the adversary is given user access on a predetermined host (in Subnet 1) that cannot be `restored' to a benign state by the defender.

\begin{figure}[b]
    \centering
    \includegraphics[width=0.6\textwidth]{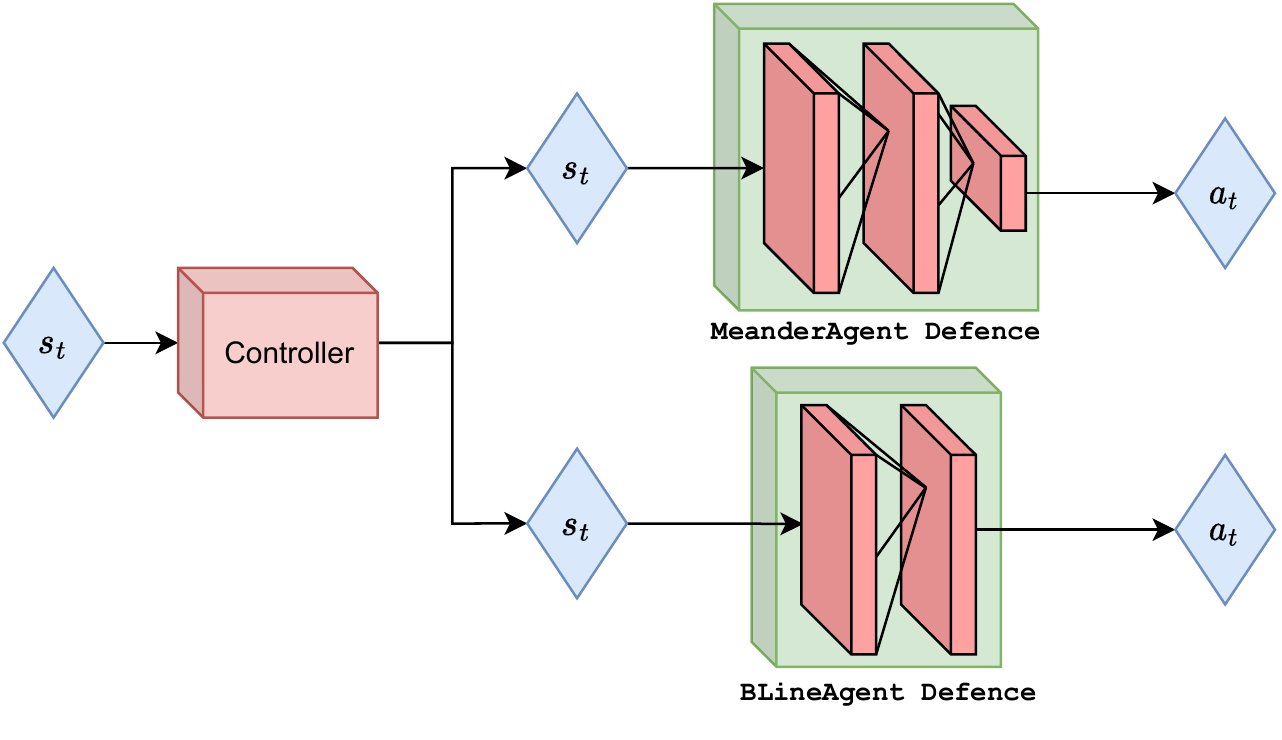}
    \caption{Hierarchical structure of the overall defensive model including the specialist subagents.}
    \label{fig:hier}
\end{figure}

\section{Model}
The models that we train have a similar basic structure to those described in~\cite{foley_autonomous_2022} that were trained for CAGE I. In particular, we focus our efforts on training a hierarchy of specialised defensive agents using DRL. These agents feature a controller agent that, at each time step, chooses a subagent to perform the action. Each subagent is trained against a specific adversarial strategy.

As described in Section~\ref{ssec:adversaries}, the environment includes two adversaries. The hierarchical architecture was developed specifically to exploit this. The model supports two expert subagents that, through the controller, are `consulted' over the course of an episode (Figure~\ref{fig:hier}). This avoids the performance limitations of a single, more general agent. Given the differences in the two adversaries, each subagent requires a different neural architecture for best performance. These are described below.


\subsection{\texttt{MeanderAgent} Defence}
Our \code{MeanderAgent} defensive subagent  was trained using the PPO algorithm and utilises a comparatively deeper neural network including three hidden layers with widths 256, 256, and 52. Full details of the hyperparameters used can be found in Appendix~\ref{app:hyperparam}.

Notably, curiosity did not improve the performance. Since the \code{MeanderAgent} is explicitly designed to explore the network during its attack, the opposing defender is also be forced to explore more broadly and to employ a wider range of strategies during training. As such, it learns sufficiently general strategies without the need for curiosity.

\subsection{\texttt{BLineAgent} Defence}
In contrast to the \code{MeanderAgent}, the \code{BLineAgent} follows a near-optimal path through the network. The \code{BLineAgent} defence, therefore, is at much greater risk of overfitting during training. As a result, we found that when training defensive agents against the \code{BLineAgent}, it was beneficial to include the curiosity mechanism. In this paper we consider two subagents for \code{BLineAgent} defence: an Action Knowledge (AK) subagent, and a State Representation (SR) subagent. Both are trained using PPO with curiosity but make different modifications to the state space.

The AK subagent modifies each observation by appending a single bit indicating the success of the previous action. We find that this gives the subagent a better understanding of the defensive process and results in an improvement in performance.

Secondly the SR agent is identical to the AK subagent, but receives observations of 27 floats as opposed to 53 bits. In this state space, each host has two floats to represent the features of activity and compromise. The additional float indicates whether the previous action succeeded. Although the mean episode reward is comparable to the AK agent's mean reward, we see a notable decrease in variance.



\section{Explaining the Adversary Model}
\begin{figure}[!t]
\centering
\begin{subfigure}{.5\textwidth}
  \centering
    \includegraphics[width=0.65\textwidth]{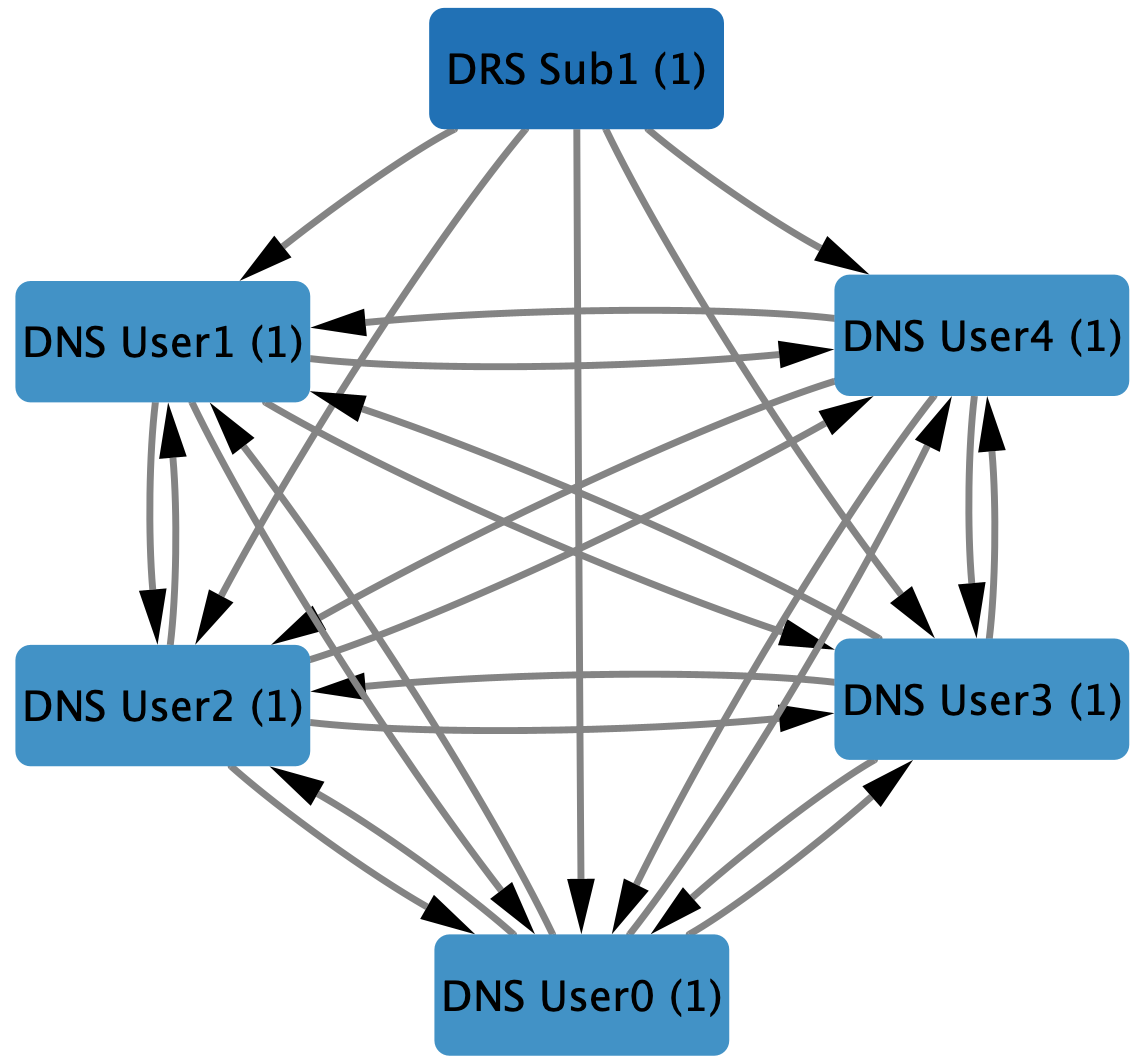}
    \caption{}
    \label{fig:meander-states_4}
\end{subfigure}%
\begin{subfigure}{.5\textwidth}
  \centering
    \includegraphics[width=0.9\textwidth]{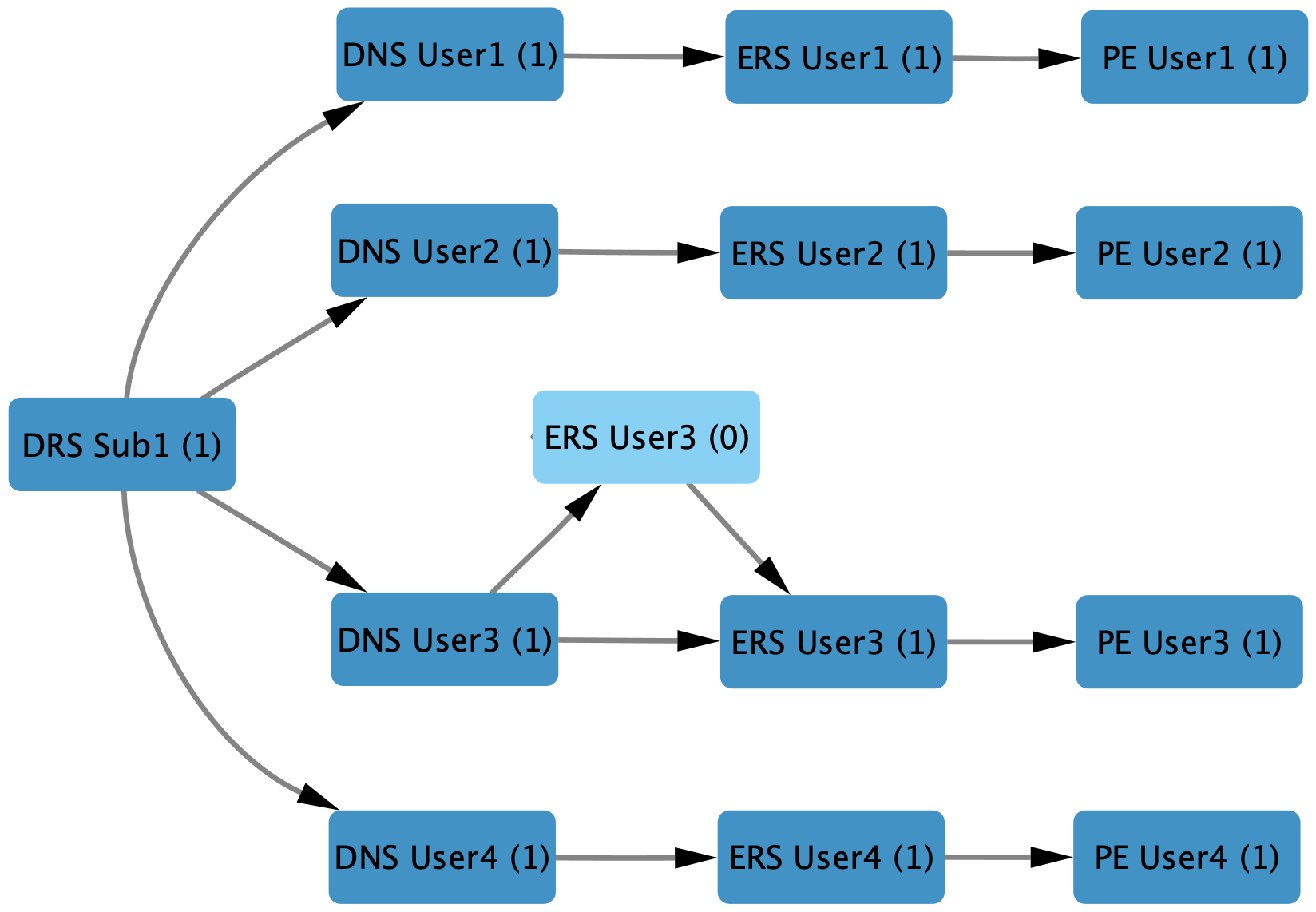}
    \caption{}
    \label{fig:bline-states_4}
\end{subfigure}
\caption{The action-outcome transition graphs of (a) the \code{MeanderAgent} and (b) the \code{BLineAgent} adversary in steps 1-4 of the CAGE II CybORG environment.}
\label{fig:explain_models}
\end{figure}

The behaviour of the adversaries is dependent on the network topology and the choice of defensive actions. In addition, there is stochasticity in both the choice and outcome of actions across all of these components. Explaining adversarial behaviour proved essential in developing effective defensive models. To better understand each adversary we, at each time step, record the choice of action, outcome and the resulting state transition. For consistency across multiple episodes we resolve IP ranges and addresses to subnets and hostnames, respectively. We observe that the connectivity (i.e., the edges) of the resulting graph provides a clear signal for differentiating the two adversaries. Figure~\ref{fig:explain_models} shows a subset of the observations, recorded during the first four steps of adversarial behaviour, in which the \code{BLineAgent} and \code{MeanderAgent} can be seen adopting a depth-first and breadth-first approach to attacking the network, respectively. In Section~\ref{sec:controller} we present two methods which make use of this observation to more accurately determine the class of adversarial threat than in prior work \cite{foley_autonomous_2022}. In Appendix~\ref{app:full-adv-modelz} we include the fully extracted adversary specifications generated by our methodology.



\section{Hierarchical RL Architecture}
\label{sec:controller}
In order to improve the performance of our defensive capability we explore the use of alternative controller models. We introduce two new types of controller for this task, one heuristic and another bandit-based.


\subsection{Bandit Controller Model}\label{ssec:bandit_controller}
We employ a bandit controller that is based on the multi-armed bandit architecture. The task is to determine which of the adversaries is currently attacking the network, based on the sequence of observations. 
However, using a bandit or bandit-like approach comes with several challenges in this setting. 

In the traditional multi-armed bandit there is no notion of state: an agent takes actions and then observes the reward. However, in the CybORG environment a unique observation cannot be used to determine the current adversary.
Thus sequences of observations need to be observed and, due to the stochasticity, there are multiple sequences that can be observed over a given number of timesteps.
A single bandit predicting the adversary will do no better than 50\%.


This is analogous to the traditional multi-armed bandit setting. Consider the task of determining which of two slot machines has the higher payout in a casino (A): the task is trivial after several attempts. Now consider a second identical casino (B) where the payout of the machines is flipped. Again, we can find the better machine in B after some error.
Finally, consider being randomly placed in A or B and having only one attempt to select the slot machine with the highest payout. As we do not know which casino we are in (as everything is identical), the best possible guess rate is 50\%.

We are able to solve this problem by abstracting the observations (which casino you are in) from the bandit. In this way we define $N_b$ bandits, one for each of the observations. As such the observation is unique to the bandit predicting the adversary. While this could also be solved by a logistic regression model, the Bandit Controller is able to learn with fewer samples, also being able to determine new adversary behaviours and learn to predict them in an online fashion. 

\subsubsection{Bandit Controller Implementation}\label{ssec:bandit_implment}

The bandit learning algorithm, shown in Algorithm \ref{alg:bandit}, allows the bandit controller to track the states that it has previously seen, creating a new bandit for each newly seen state. Each of these bandits is initialised with $Q$ values for each of the actions $a \in \{0, 1, 2\}$, where these correspond to the \code{MeanderAgent}, \code{BLineAgent}, and no adversary. The $Q$ values are updated using reward $R$ and the number of times that prediction has been selected, $N(A)$. We train the bandit controller for 15,000 timesteps, using $epsilon = 0.01$.

The Bandit Controller has a state different to that of its subagents. Its state is a sliding window of the last four timesteps from the CybORG environment. As we can see from Figure~\ref{fig:bline-states_4}, the minimum number of actions before an adversary has user privilege (and the first unambiguous instance of malicious behaviour) is three. A defensive agent can observe this on the fourth timestep, hence a prediction from the bandit controller only needs to happen once per episode. Finally, we use a simple reward function of +1 for a correct prediction, and -1 for an incorrect prediction. 

 \begin{algorithm}[h]
\DontPrintSemicolon
  \SetStartEndCondition{ }{}{}%
\SetKwProg{Fn}{\ttfamily\textcolor{blue}{Function}}{\string:}{}
\SetKwFunction{Range}{range}
\SetKw{KwTo}{in}\SetKwFor{For}{\ttfamily\textcolor{blue}{for}}{\string:}{}%
\SetKw{KwTo}{in}\SetKwFor{Initialise}{\ttfamily\textcolor{blue}{Initialise}}{\string:}{}%
\SetKw{KwTo}{in}\SetKwFor{Predict}{\texttt{\textcolor{blue}{Predict}}($s$)}{\string:}{}%
\SetKwIF{If}{ElseIf}{Else}{\ttfamily\textcolor{blue}{if}}{:}{\ttfamily\textcolor{blue}{elif}}{\ttfamily\textcolor{blue}{else:}}{}%
  {
    \texttt{\textcolor{blue}{Initialise}} the known states, $s_n$
    
    \texttt{\textcolor{blue}{Initialise}} set of bandits, $B$ 

    \Initialise{for \textit{a} = 1 to \textit{k}}{
         $bandit_0.Q(a) \leftarrow 0$        \tcp{Initialise Q values and action counter for the first bandit}

         $bandit_0.N(a) \leftarrow 0$
    }
    \Predict{}{
    \If{$s \not\in  s_n$}{
        $s_n$ $\leftarrow$ $s$
        
        \textcolor{blue}{\texttt{Initialise}} $bandit_s$
        
        $B \leftarrow bandit_s$
    }
    
    $A \leftarrow$
    $\begin{cases}
                    argmax_a(bandit_s.Q(a)) & \text{with probability 1 - $\epsilon$}  \\
                    random\ action & \text{with probability $\epsilon$}
    \end{cases}$
    
    $R \leftarrow prediction\_result(A)$
    
    $bandit_s.N(A) \leftarrow bandit_s.N(A) + 1$
    
    $bandit_s.Q(A) \leftarrow bandit_s.Q(A) + \frac{1}{N(A)}\big[R-bandit_s.Q(A)]$
    
    }
}

\caption{Bandit Controller Learning Algorithm.}
\label{alg:bandit}
\end{algorithm}

\subsection{Heuristic Controller Model}
\label{ssec:heuristic_controller}

We also construct a heuristic for predicting the adversary. This approach is possible as we are able to observe the patterns that the adversaries display in a controlled version of the CybORG environment. As we can see in Figures \ref{fig:bline-states_4} and \ref{fig:meander-states_4}, the \code{BLineAgent} and \code{MeanderAgent} have fundamentally different strategies in the first four moves they make. Using this privileged view of the adversarial behaviour allows for a manual and formal definition of the behaviour, as defined in Heuristic~\ref{heuristic:controller}. As in the Bandit Controller we use this heuristic once per episode, on the fourth timestep, to determine which adversary is attacking the network.

\begin{heuristic}\label{heuristic:controller}
The scanning of two different hosts on the network within the first four timesteps indicates the presence of the \code{MeanderAgent} adversary. Otherwise, this is either the \code{BLineAgent} adversary or the \code{User} agent.
\end{heuristic}



\section{Evaluation}\label{sec:evaluation}

In this section we evaluate the performance of our specialist subagents against the two adversaries. We further investigate the performance of the controller models. Finally, we evaluate the full defensive model capable of defending against either adversary. 
We use the model described in prior work~\cite{foley_autonomous_2022} as a baseline performance measure (baseline for brevity), as this has been established as state-of-the-art and achieved the best score in CAGE I. Because the scoring function assigns only penalty points (i.e., 0 is the theoretically maximum score), all the reported rewards are negative.


\begin{figure}[!t]
\centering
\begin{minipage}{0.998\textwidth}

\begin{subfigure}{.49\textwidth}
  \centering
  \includegraphics[width=.9\linewidth]{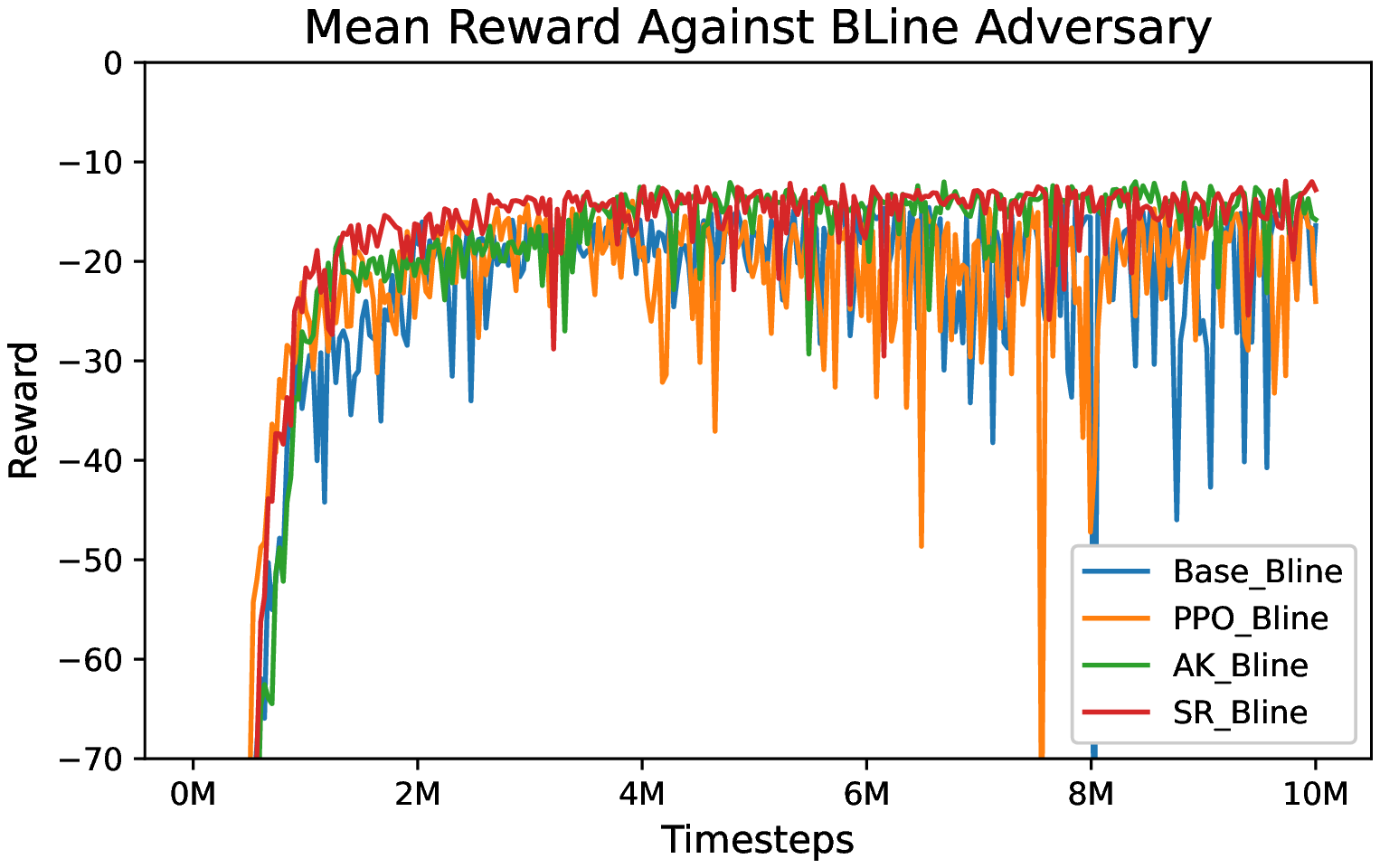}
  \caption{}
  \label{fig:sub1}
\end{subfigure}%
\begin{subfigure}{.49\textwidth}
  \centering
  \includegraphics[width=.9\linewidth]{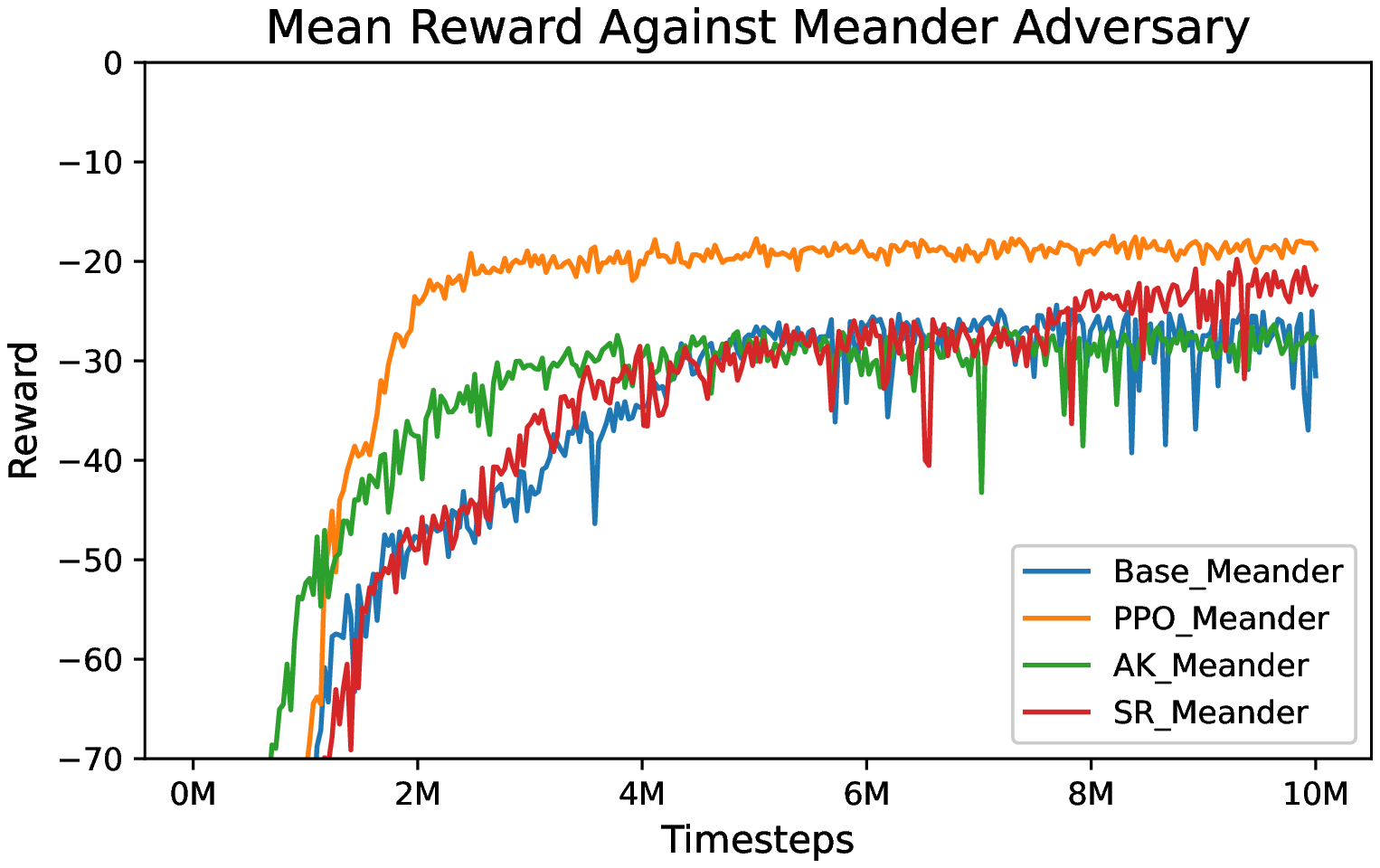}
  \caption{}
  \label{fig:sub2}
\end{subfigure}
\begin{subfigure}{.49\textwidth}
  \centering
  \includegraphics[width=.9\linewidth]{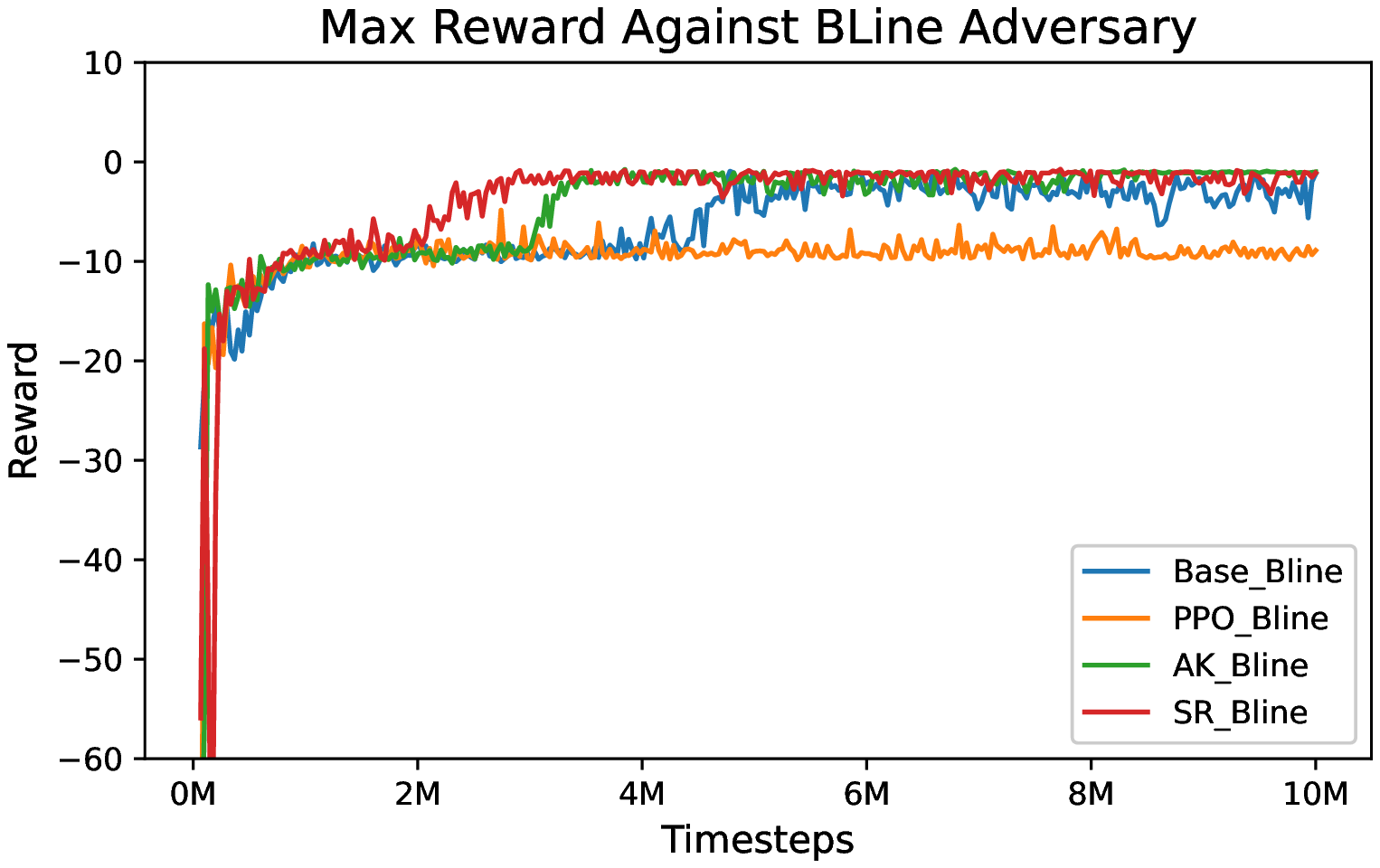}
  \caption{}
  \label{fig:sub3}
\end{subfigure}%
\begin{subfigure}{.49\textwidth}
  \centering
  \includegraphics[width=.9\linewidth]{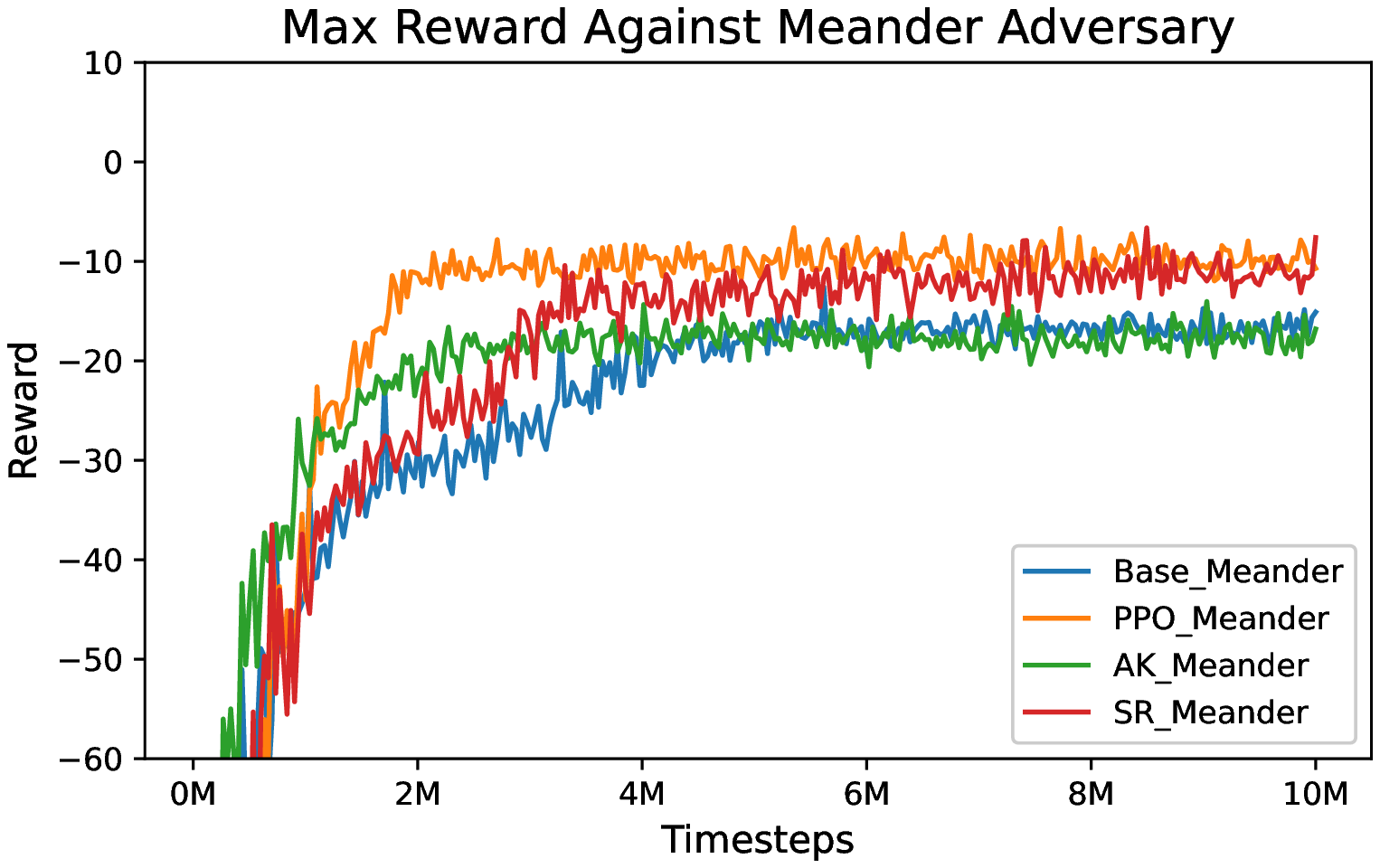}
  \caption{}
  \label{fig:sub4}
\end{subfigure}
\begin{subfigure}{.49\textwidth}
  \centering
  \includegraphics[width=.9\linewidth]{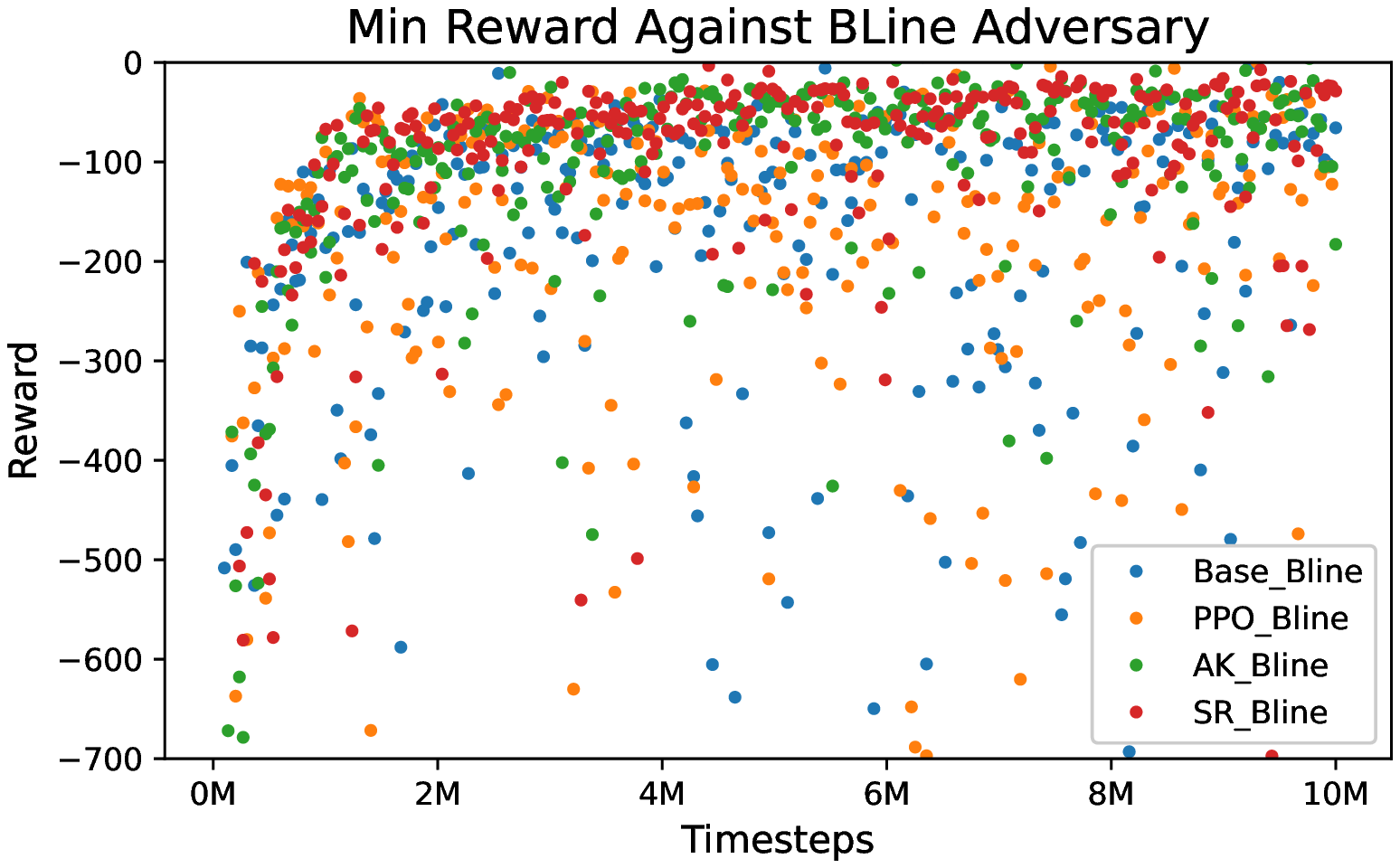}
  \caption{}
  \label{fig:sub5}
\end{subfigure}%
\begin{subfigure}{.49\textwidth}
  \centering
  \includegraphics[width=.9\linewidth]{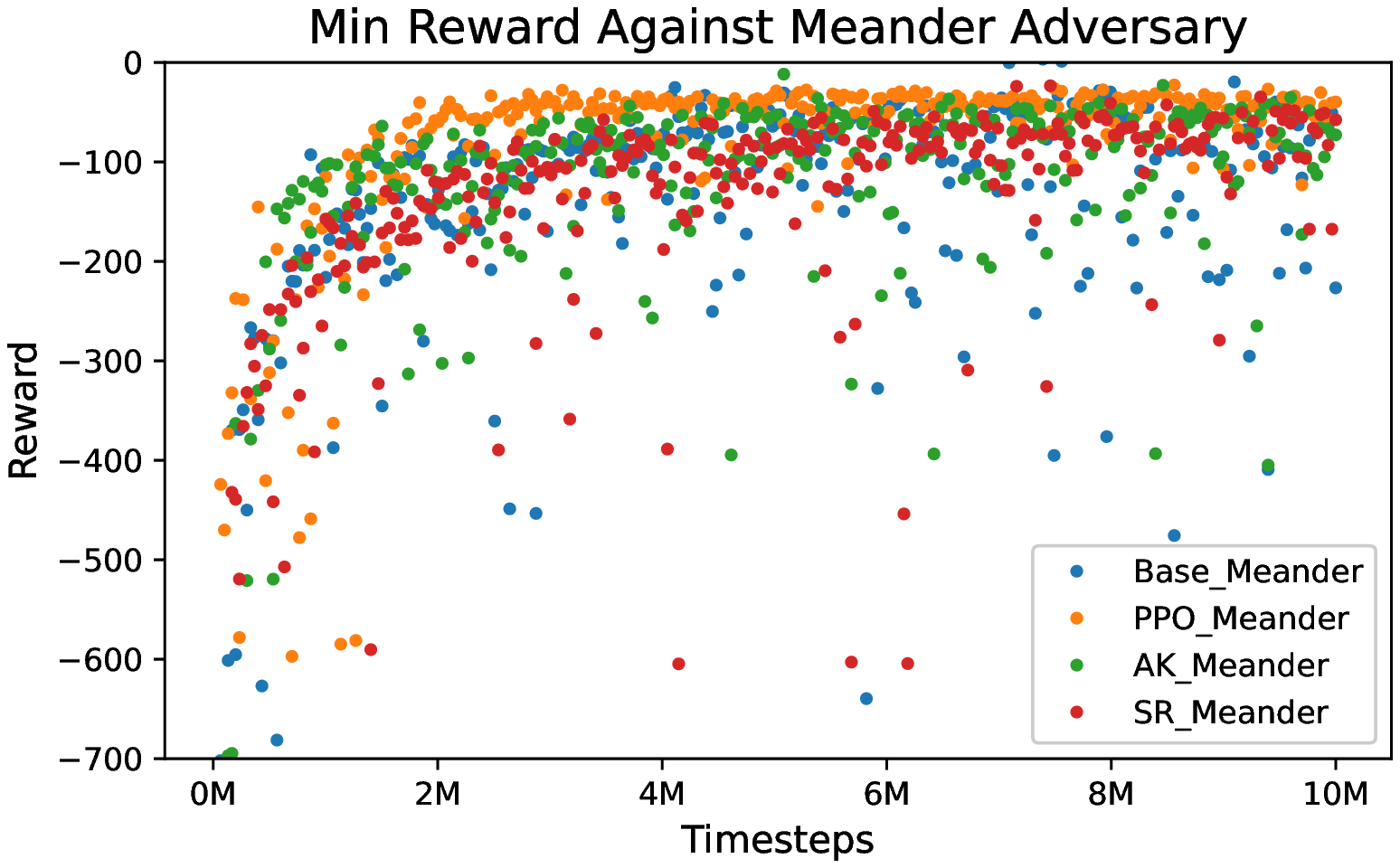}
  \caption{}
  \label{fig:sub6}
\end{subfigure}
\begin{subfigure}{.49\textwidth}
  \centering
  \includegraphics[width=.9\linewidth]{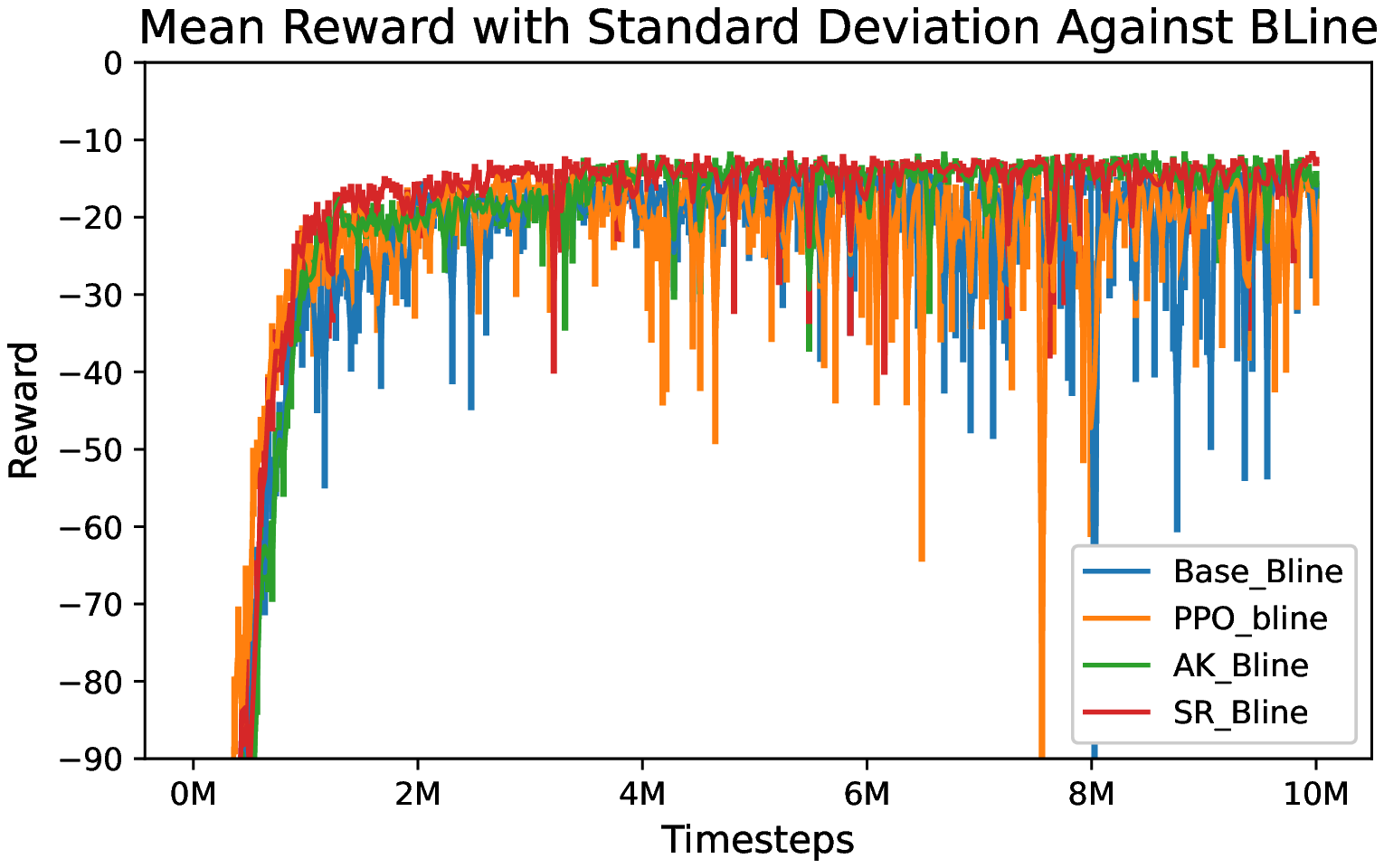}
  \caption{}
  \label{fig:sub7}
\end{subfigure}%
\begin{subfigure}{.49\textwidth}
  \centering
  \includegraphics[width=.9\linewidth]{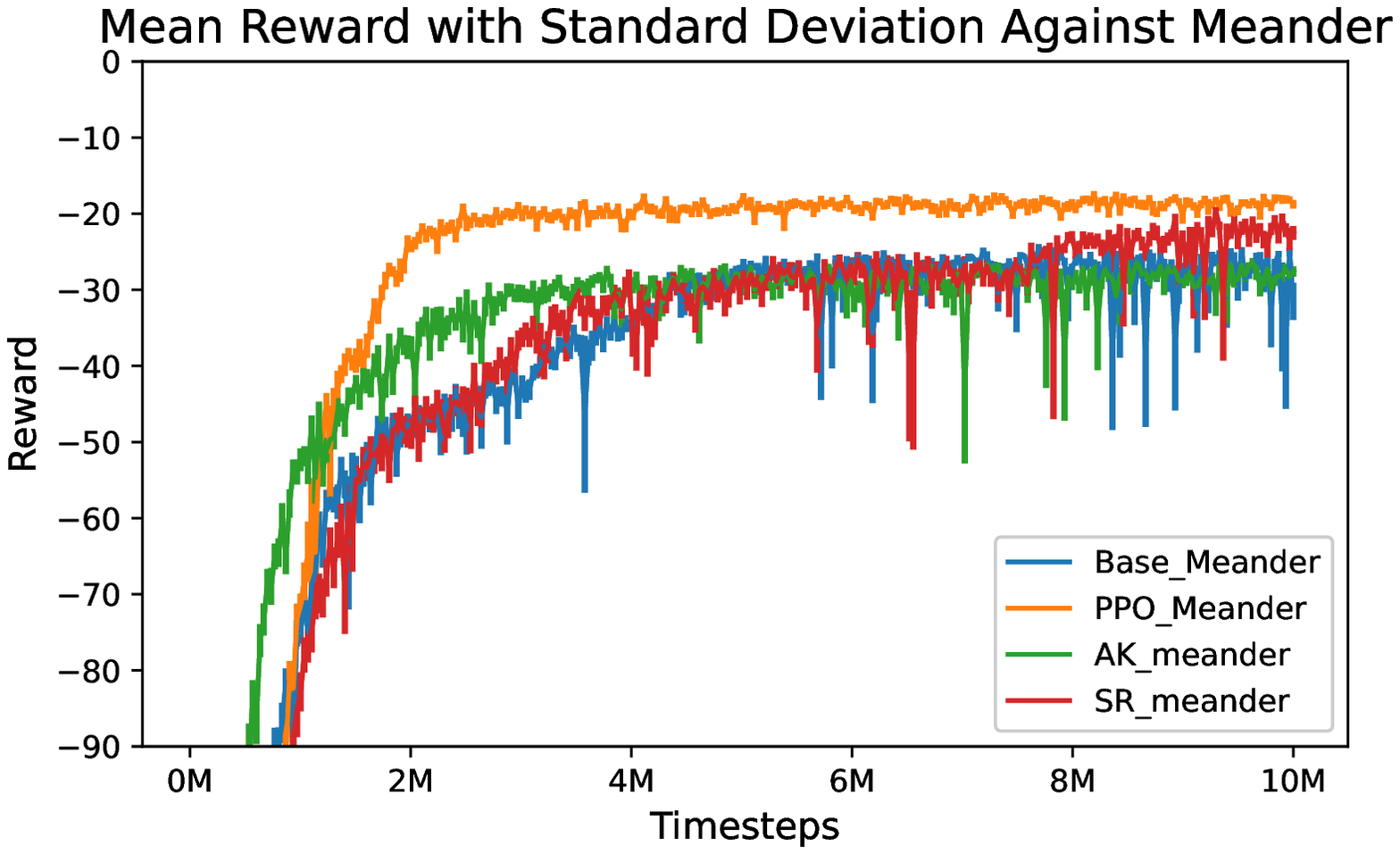}
  \caption{}
  \label{fig:sub8}
\end{subfigure}
\end{minipage}
\caption{Mean, maximum and minimum reward of blue subagents against the \code{BLineAgent} (left) and the \code{RedMeanerAgent} (right) over 10 million timesteps.}
\label{fig:sub_agent_training}
\end{figure}

\subsection{Specialised Sub Agents}\label{ssec:eval_sub}


\subsubsection{Training Results}
Figure~\ref{fig:sub_agent_training} shows the average reward of each defensive subagent as trained against the \code{BLineAgent} (left column), and the \code{MeanderAgent} (right column).
The methods of AK and SR achieve peak rewards against the \code{BLineAgent} of -12.227 and -11.465 respectively, both of which are an improvement over the baseline~\cite{foley_autonomous_2022} PPO with curiosity based model, which achieves -13.475. Furthermore, removing curiosity negatively impacts the reward against the \code{BLineAgent}, as shown clearly in the max reward plot of Figure~\ref{fig:sub_agent_training}(c) 

The difference in mean reward is explained by the maximum and minimum rewards. All models apart from PPO experience a first plateau in maximum reward of -9 and then step up to a second plateau of around -1. The SR agent finds the optimal policy earlier than the AK agent during training. In addition, the minimum rewards of the baseline and PPO model have greater variance than the SR and AK agents, and the AK agent has a marginally higher probability to score very poorly (i.e., below -300). Earlier optimal policy convergence and smaller policy variability makes the SR agent the best model against \code{BLineAgent} agent. This corroborates the standard deviation graph, and 1,000-episode evaluation results; in Table~\ref{table:sunagent_perf}, where the SR agent displays less negative reward and with a standard deviation that is only a fifth of the AK agent. 

Against the meander attacker the PPO and SR agents outperform the baseline (-24.384) with best mean rewards of -17.065 and -19.959 during training. Figure~\ref{fig:sub_agent_training} shows the advantage of using a PPO 3-layer architecture which results in higher min and max rewards with reduced variance.

\subsubsection{Specialist Agents}\label{ssec:specialists}
Here we evaluate the performance of our defensive subagents against their separate adversaries. We select the best performing agents from training for evaluation evaluate: PPO defence for the \code{RedMeander} and both the AK and SR defence for the \code{BLineAgent}. We evaluate each for 1,000 episodes of 100 steps and summarise our results in Table~\ref{table:sunagent_perf}. For completeness, we also cross-evaluate our agents against the adversary not seen during training.

Against the \code{RedMeander} adversary, PPO defence outperforms the baseline against both adversaries resulting in a mean score of -21.3 (improvement by a factor of 3.6) and a reduction in standard deviation by a factor of more than 9. This highlights the advantage of the increased depth of the neural network over the baseline. 

Against the \code{BLineAgent} adversary, we see that the SR agent is able to achieve a 1.5 times greater reward, with 4.89 times lower standard deviation. However, this comes at the cost of generality. A trend in all of the subagents is that when defending against previously unseen adversaries, the performance is significantly diminished. 



\begin{table}[]
\centering
{\footnotesize
\begin{tabular}{@{}llccccll@{}}
\toprule
\multirow{2}{*}{\begin{tabular}[c]{@{}l@{}}Training\\ Adversary\end{tabular}}              & \multirow{2}{*}{Defensive Model} & \multicolumn{2}{c}{\texttt{BLineAgent}}                                                                                                                       & \multicolumn{2}{c}{\texttt{MeanderAgent}}                                                                                                                     & \multicolumn{2}{c}{Mean}                                                                                              \\ \cmidrule(l){3-8} 
 &                                  & \multicolumn{1}{l}{\begin{tabular}[c]{@{}l@{}}Mean \\ Reward\end{tabular}} & \multicolumn{1}{l}{\begin{tabular}[c]{@{}l@{}}Standard\\ Deviation\end{tabular}} & \multicolumn{1}{l}{\begin{tabular}[c]{@{}l@{}}Mean \\ Reward\end{tabular}} & \multicolumn{1}{l}{\begin{tabular}[c]{@{}l@{}}Standard \\ Deviation\end{tabular}} & \begin{tabular}[c]{@{}l@{}}Mean \\ Reward\end{tabular} & \begin{tabular}[c]{@{}l@{}}Standard\\  Deviation\end{tabular} \\ \midrule
\texttt{MeanderAgent}  & PPO    &  \textbf{-24.91}  & \textbf{9.21}  &   \textbf{-17.71}   &   \textbf{5.06}       &  \textbf{-21.31}   &  \textbf{7.43}  \\ 
                & baseline     & -123.91    & 229.59     &  -30.39    & 47.74   &   -77.15  &      165.82    \\ \hline
\multirow{3}{*}{\texttt{BLineAgent}} & AK    & -14.43   & 30.28     &  \textbf{ -145.26}   &  \textbf{123.92}     &    \textbf{-79.84}  &    \textbf{90.20}     \\ 
                                          & SR     & \textbf{-12.95}    & \textbf{6.19}     &  -269.64   &    235.24    &   -139.29    &   166.40  \\ 
                                          & baseline    & -16.80  & 21.12    &   -201.13   &  143.40  &   -108.96   &      102.49       \\ \bottomrule
                                          
\end{tabular}}
\vspace{+2mm}
\caption{The performance of the defensive subagents against their corresponding adversaries. Evaluated on 1,000 episodes of 100 timesteps each.}\label{table:sunagent_perf}

\end{table}

\subsection{Controller Models}\label{ssec:eval_control}
As seen in Section~\ref{ssec:specialists}, the defensive subagents do not generalise well beyond the adversaries that they are trained against.

To address this, Sections~\ref{ssec:heuristic_controller} and~\ref{ssec:bandit_controller}
introduce two new controller architectures: Heuristic and Bandit. Here we evaluate the ability to correctly predict the adversary within the first four timesteps of an episode (as our controllers predict the adversary on the fourth timestep). For each episode, we randomly sampled one of the two red adversaries (i.e., 50\% probability of selecting \code{BLineAgent}).


Table~\ref{table:controller_pred} shows that the baseline model has strong biases on selecting the \code{BLineAgent} agent. To investigate further, we let the baseline agent make predictions on each timestep until the end of the episode (c.f. only guessing after the 4th timestep). As seen in Table~\ref{table:controller_pred}, the repeated guesses significantly reduced bias but accuracy remained low.

In contrast, neither our bandit or heuristic controller exhibit this bias and can perfectly predict the correct attacker type. 


\begin{table}[]
\centering
\begin{tabular}{@{}l|cc@{}}
\toprule
\multicolumn{1}{c|}{\multirow{2}{*}{Controller Agent}} & \multicolumn{2}{c}{Prediction Accuracy} \\ \cmidrule(l){2-3} 
\multicolumn{1}{c|}{}                                  & \texttt{BLineAgent}   & \texttt{RedMeander}   \\ \midrule
PPO with curiosity (4 steps)                           & 76.8\%              & 0.0\%               \\
PPO with curiosity (100 steps)                         & 30.3\%              & 42.9\%              \\
Heuristic                                              & 100.0\%             & 100.0\%             \\
Bandit                                                 & 100.0\%             & 100.0\%             \\ \bottomrule
\end{tabular}
\vspace{+2mm}
\caption{Controller performance taken over 1,000 episodes of 4 steps, except in the case of PPO with curiosity (100 steps) which predicts the adversary at each timestep.}\label{table:controller_pred}
\end{table}


\subsection{Hierarchical Defensive Model}\label{ssec:eval_complete}
 Here we evaluate the complete defensive model. Table~\ref{table:overall_eval} reports the mean and standard deviation for the `best pair' combinations of subagents as determined by our evaluation in Section~\ref{ssec:eval_sub} ( i.e., PPO for \code{MeanderAgent}, and AK or SR for \code{BLineAgent}).

We observe that the subagents play a significant role in the improvement over the baseline. Over episodes of 100 timesteps, we are able to improve the result by at least 30\% for the \code{BLineAgent} and 170\% for \code{MeanderAgent}. 

The lowest reward values are split evenly between the Heuristic and Bandit controllers. These models outperform the PPO controller models regardless
of the subagents in four of the six combinations of adversary and episode length. 

\code{MeanderAgent} performance is improved by 11.7\%, which is more significant than \code{BLineAgent} (only improved by 1\%) when using Bandit or Heuristic controller. Table~\ref{table:sunagent_perf} indicates that models trained with \code{BLineAgent} perform poorly on \code{MeanderAgent}. This can be explained by the fact that \code{BLineAgent} has more information about the network, so its behaviour is more predictable. In contrast, \code{MeanderAgent}'s actions have more randomness.


\begin{table}[]
\centering{\fontsize{6.7pt}{8pt}\selectfont
\begin{tabular}{@{}p{1.8cm}p{1.7cm}cccccc@{}}
\toprule
\multirow{2}{*}{Controller}              & \multirow{2}{*}{Subagents} & \multicolumn{2}{c}{\texttt{30  steps}}                                                                                                                       & \multicolumn{2}{c}{\texttt{50 steps}}                                                                                                                     & \multicolumn{2}{c}{100 steps}                                                                                              \\ \cmidrule(l){3-8} 
 &                                  & \texttt{\fontsize{6.7pt}{8pt}\selectfont BLineAgent} & \multicolumn{1}{c}{\texttt{\fontsize{6.7pt}{8pt}\selectfont MeanderAgent}} & \multicolumn{1}{c}{\texttt{\fontsize{6.7pt}{8pt}\selectfont BLineAgent}} & \multicolumn{1}{c}{\texttt{\fontsize{6.7pt}{8pt}\selectfont MeanderAgent}} & \texttt{\fontsize{6.7pt}{8pt}\selectfont BLineAgent} & \texttt{\fontsize{6.7pt}{8pt}\selectfont MeanderAgent} \\ \midrule
\multirow{2}{*}{Bandit}  & PPO + AK    & \textbf{-3.56$\pm$2.03} &  \textbf{-6.80$\pm$1.40}   &   -6.79$\pm$13.00      &  -10.10$\pm$2.30   &   -13.54$\pm$15.95 & \textbf{-17.30$\pm$4.27}\\ 
                & PPO + SR     &  -3.62$\pm$2.04    & -6.88$\pm$1.42     &   -6.26$\pm$3.18   & -10.06$\pm$2.15   &  \textbf{-13.00$\pm$6.28}   &    -17.56$\pm$4.51      \\ \midrule
\multirow{2}{*}{Heuristic}  & PPO + AK   &  -3.56$\pm$2.04  & \textbf{-6.80$\pm$1.40 }&    -6.79$\pm$13.00   &  \textbf{ -9.96$\pm$2.33 }  &  -14.07$\pm$27.73   &  -17.57$\pm$4.82  \\ 
                & PPO + SR     & -3.71$\pm$2.09    & -6.86$\pm$1.48    & \textbf{ -6.17$\pm$3.40}    & -10.04$\pm$2.32   &  -13.06$\pm$6.14   &    -17.32$\pm$4.35      \\ \midrule
\multirow{2}{*}{\begin{tabular}[c]{@{}l@{}} Baseline  \\ (PPO Controller)\end{tabular}} & PPO + AK  &  -4.35$\pm$2.42 & -7.19$\pm$1.69   &  -7.45$\pm$4.27   &   -10.84$\pm$2.62    &  -14.97$\pm$8.09   &    -19.33$\pm$5.38     \\ 
                    & PPO + SR    & -3.95$\pm$2.18  &  -7.36$\pm$1.74   &  -6.38$\pm$3.20   &  -11.33$\pm$3.00  &   -13.14$\pm$6.45  & -21.21$\pm$6.10            \\ \midrule
\multirow{2}{*}{\begin{tabular}[c]{@{}l@{}} Baseline  \\ (PPO Controller)\end{tabular}} & \multirow{2}{*}{\begin{tabular}[c]{@{}l@{}} Baseline  \\ (PPO subagents)\end{tabular}} &   4.82$\pm$4.22 & -8.78$\pm$3.21   & -9.20$\pm$16.01     &   -19.00$\pm$20.86    &    -18.49$\pm$34.40 &  -47.60$\pm$88.16       \\ 
\\
                                            
                    \bottomrule
                                          
\end{tabular}}
\vspace{+2mm}
\caption{Performance of all subagents-controller combinations, evaluated over 1,000 episodes with a length of 30, 50 and 100 steps each.}\label{table:overall_eval}

\end{table}

\section{Explaining the Defensive Models}\label{sec:explain_defence}


It is critically important that human operators can understand the decisions made by autonomous agents. Using post-hoc XRL techniques, we determine whether our defensive agents are truly defending the network as their primary objective or as a side effect of an unintended objective. This is common in RL where agents may manipulate improperly specified reward mechanics to maximise their score in unintended ways. 


\subsection{Ablation Study}\label{ssec:ablation}

To understand which of the features in the observation space influence the agents decision making we perform an ablation study over knowledge of: 1) the success or failure of the previous action (hence referred to as \emph{previous action}), 2) the adversary's access onto a host (hence referred to as \emph{adversary access}), and 3) whether an adversary has scanned a host (hence referred to as \emph{adversary scan}). 

\begin{figure}[!t]
\centering
\begin{subfigure}{.33\textwidth}
  \centering
  \includegraphics[width=\linewidth]{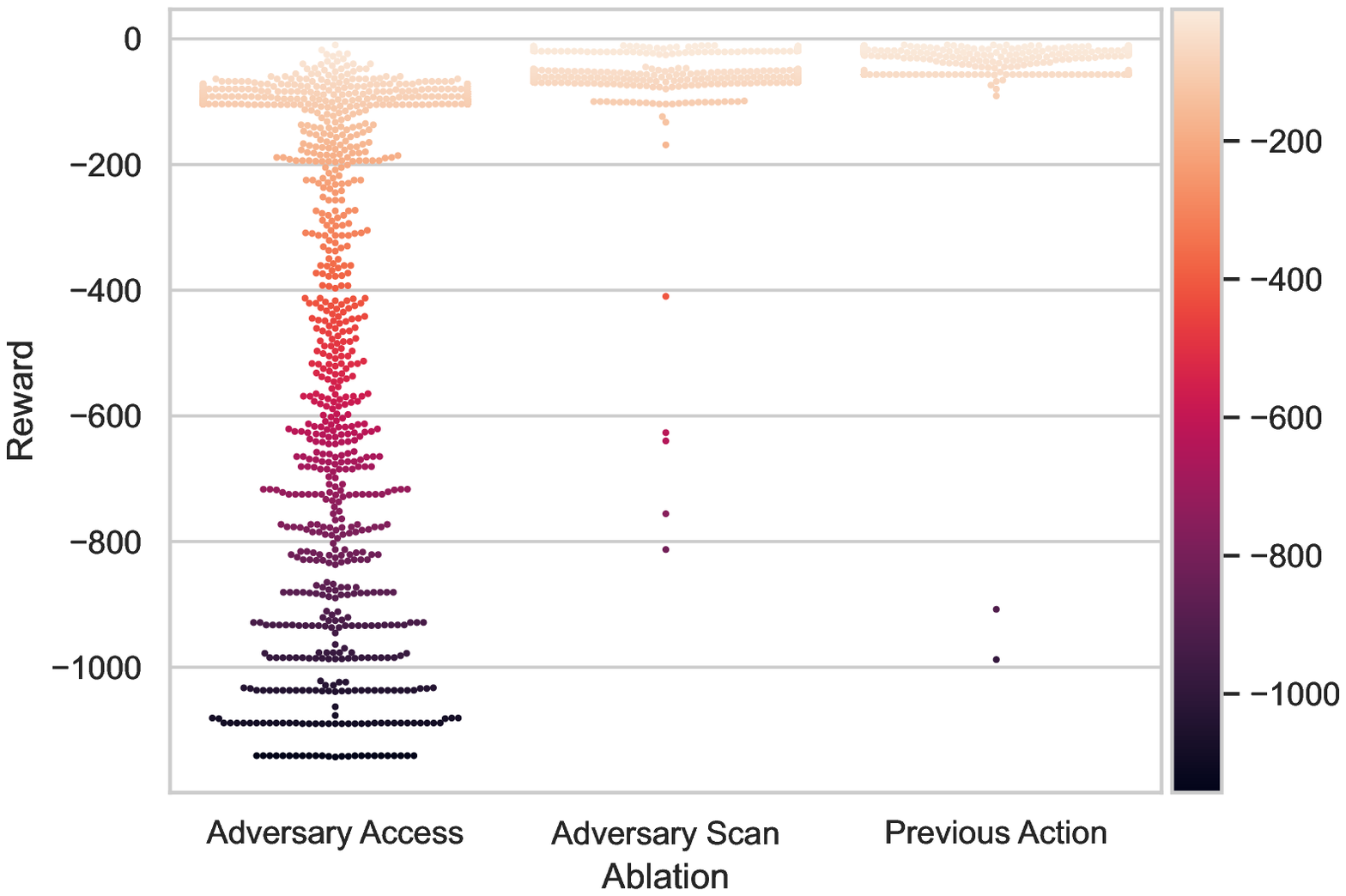}
  \caption{}
  \label{fig:ablation_bline}
\end{subfigure}%
\begin{subfigure}{.33\textwidth}
  \centering
  \includegraphics[width=\linewidth]{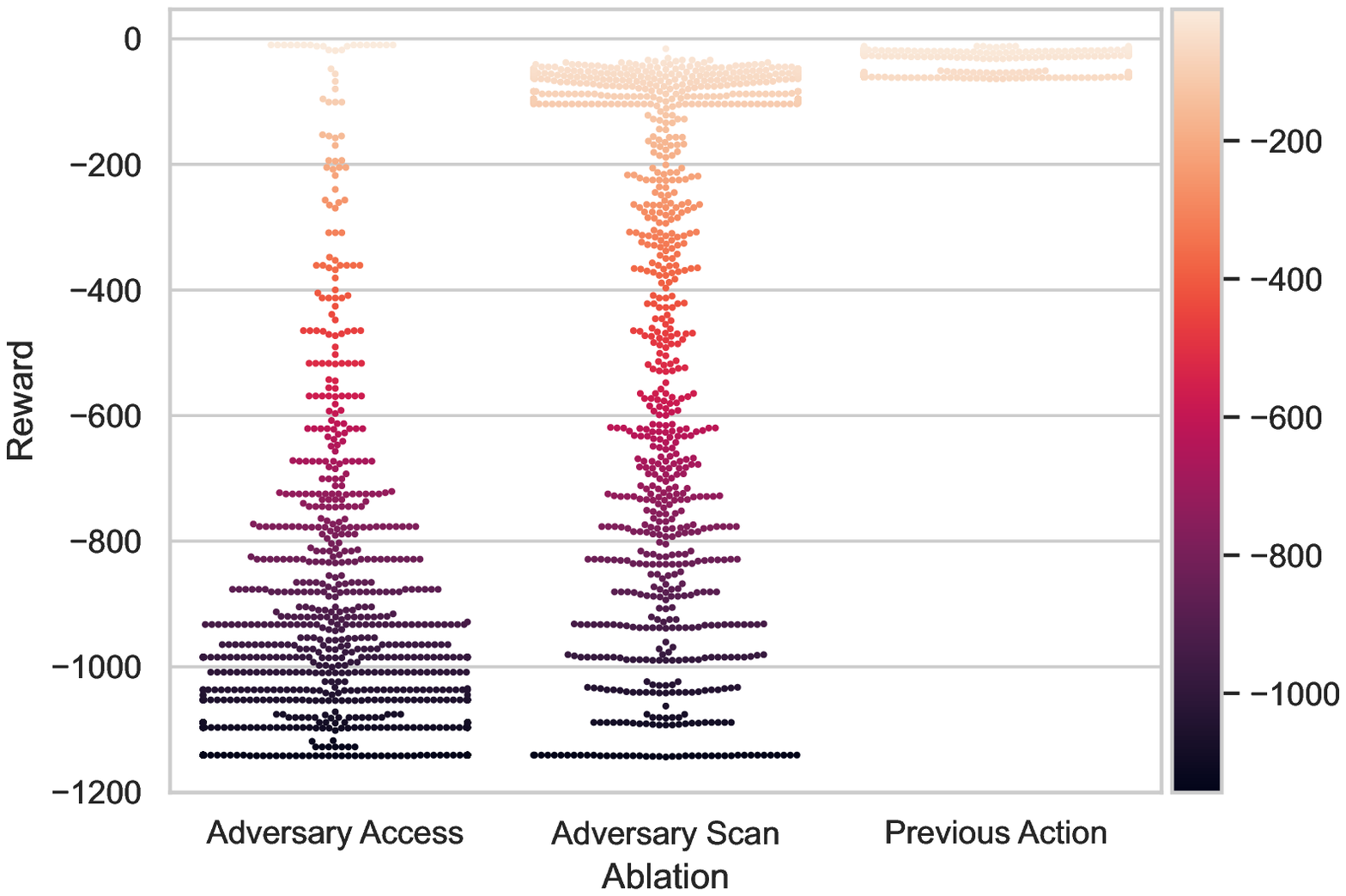}
  \caption{}
  \label{fig:ablation_bline_flaor}
\end{subfigure}%
\begin{subfigure}{.33\textwidth}
  \centering
  \includegraphics[width=\linewidth]{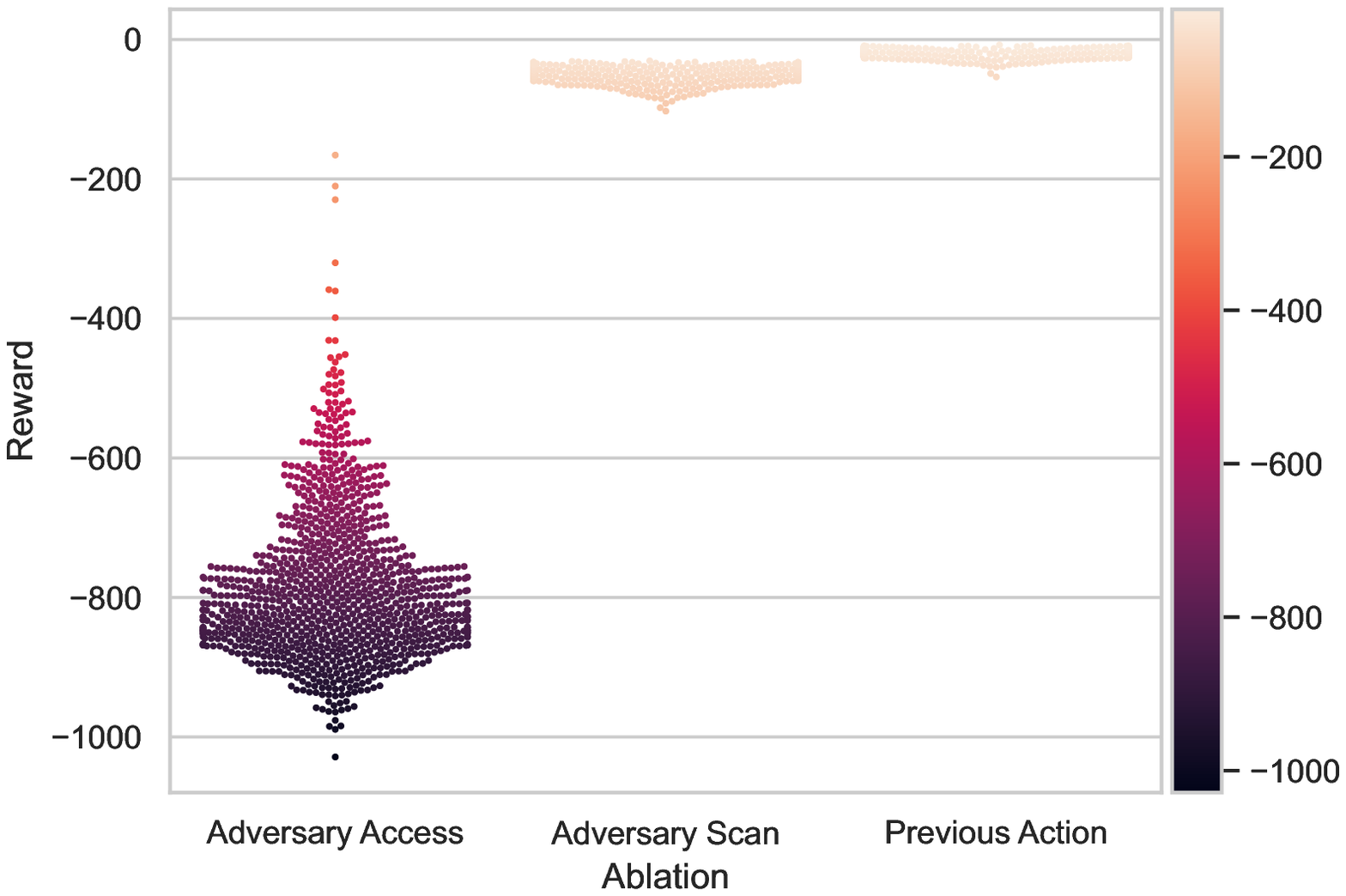}
  \caption{}
  \label{fig:ablation_meander}
\end{subfigure}
\caption{Ablation results for the three feature types on the reward of the (a) AK \texttt{BLineAgent} defence, (b) SR \texttt{BLineAgent} defence  and (c) PPO \texttt{MeanderAgent} defence.}
\label{fig:ablation}
\end{figure}

The ablation results in Figure~\ref{fig:ablation} show the AK and SR agents against the \code{BLineAgent} in \ref{fig:ablation_bline} and~\ref{fig:ablation_bline_flaor}, and the PPO agent against the \code{MeanderAgent}\footnote{This defensive agent doesn't use the previous action however we include it for completeness.} in \ref{fig:ablation_meander}.  Figure~\ref{fig:ablation_bline} indicates that the AK agent's performance is greatly affected by `adversary access'. While comparatively little impact seems to derive from the ablation of `adversary scan' and `previous action' there is some variance and the rewards fall to -812 and -539, respectively. Interestingly, the SR defensive agent is greatly affected by the ablation of the `adversary access' and `adversary scan', with the distribution of rewards being more negative in both cases. This is especially apparent in the case of `adversary scan'. Previous action has less of an effect in both AK and SR, yet still reduces the mean reward to -30.42 (a factor of 2) and -40.6 (a factor of 3), respectively. However, AK has some outlier scores that result in a minimum reward of -987.8. For PPO against \code{MeanderAgent}, Figure~\ref{fig:ablation_meander} shows that ablation of `adversary access' causes a drastic reduction in reward, bringing the mean value to -781.23. Ablation of `adversary scan' reduces the mean reward to -44.70, a factor of 2.52 more negative than when the observation is included. 

\subsection{Feature Importance}\label{ssec:shap}

To further validate the importance of `adversary access', `previous action', and `adversary scan', we utilise a well known framework from explainable AI called SHapley Additive exPlanations (SHAP). This uses an implementation agnostic game theoretic approach to explain the importance of features in determining outputs. SHAP is able to connect optimal credit allocations with local explanations to determine SHAPley values. These values provide a way of accurately distributing the contribution of the individual features within the complete feature space~\cite{lundberg_unified_2017}.

\begin{figure}[!h]
\centering
\begin{subfigure}{.5\textwidth}
  \centering
  \includegraphics[width=\linewidth]{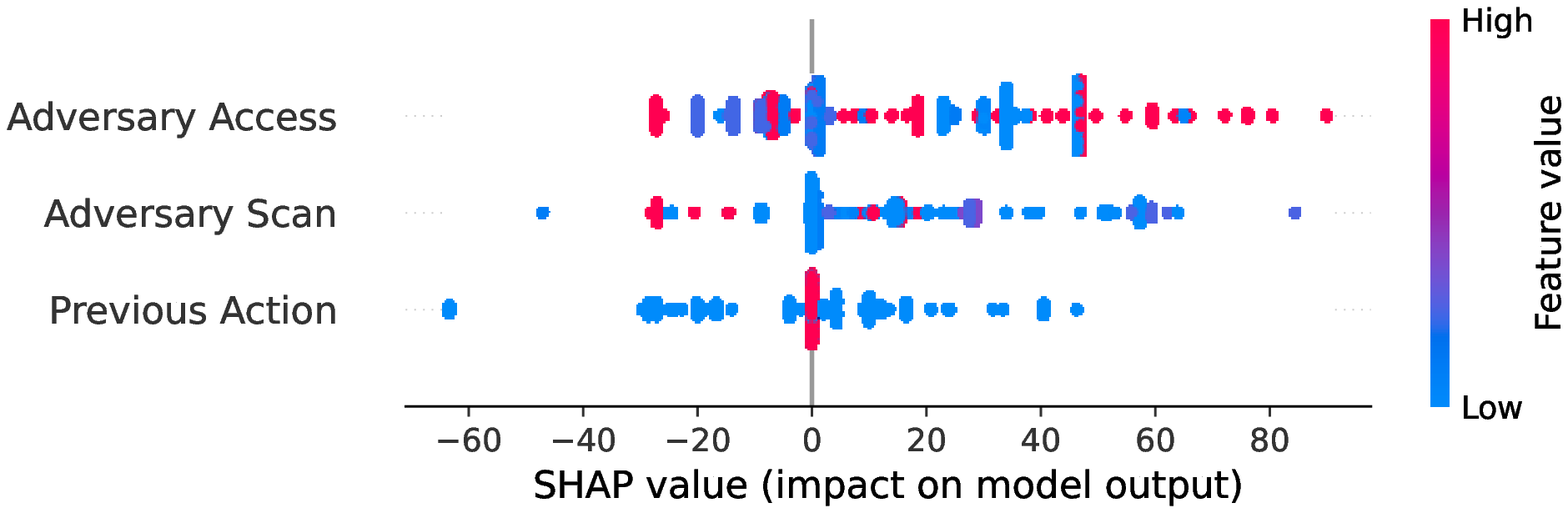}
  \caption{}
  \label{fig:shap_bline}
\end{subfigure}%
\begin{subfigure}{.5\textwidth}
  \centering
  \includegraphics[width=\linewidth]{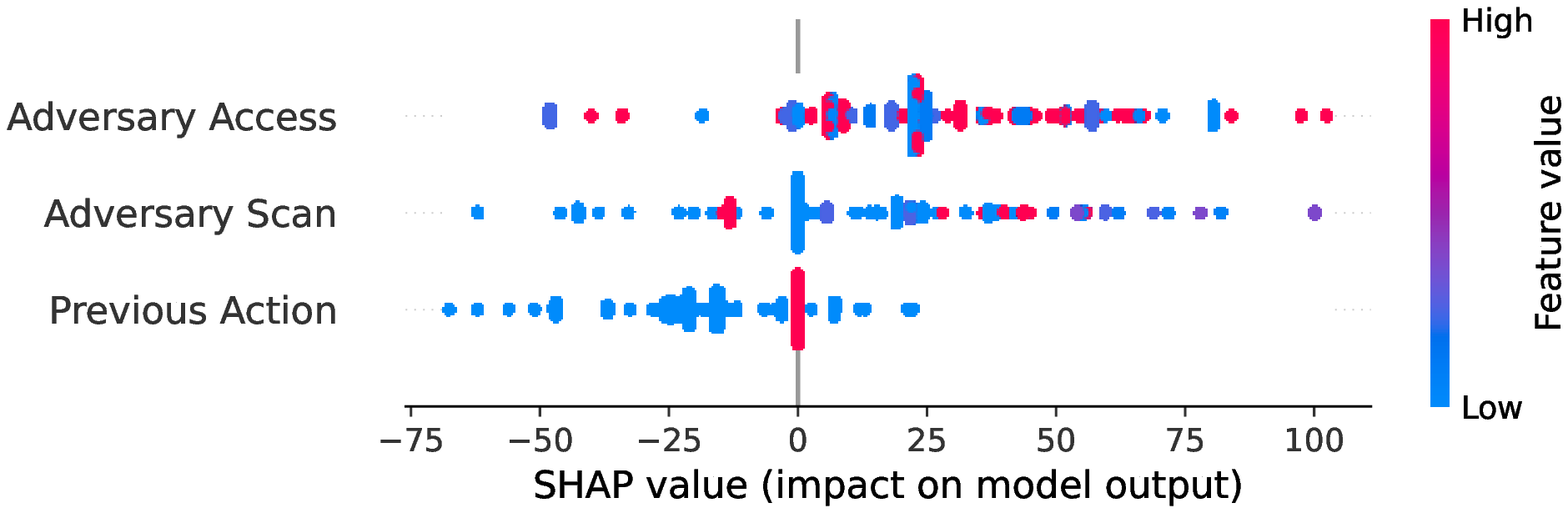}
  \caption{}
  \label{fig:shap_bline_float}
\end{subfigure}

\begin{subfigure}{.5\textwidth}
  \centering
  \includegraphics[width=\linewidth]{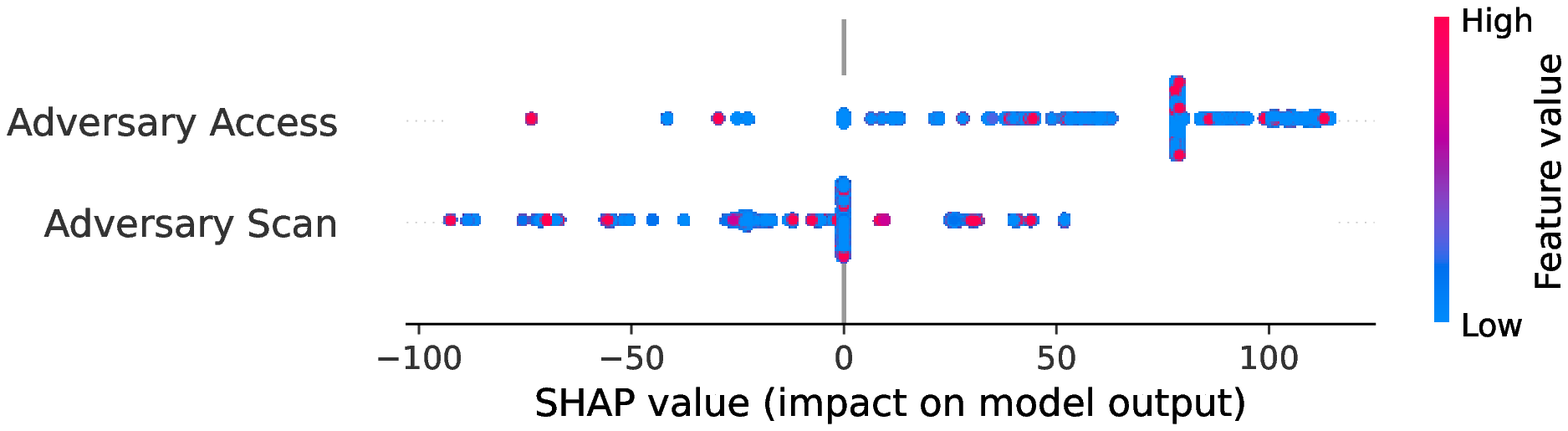}
  \caption{}
  \label{fig:shap_meander}
\end{subfigure}
\caption{The SHAPley values for the three feature types on the reward of the (a) AK \texttt{BLineAgent} defence, (b) SR \texttt{BLineAgent} defence and (c) PPO \texttt{MeanderAgent} defence, from most to least important features.}
\label{fig:shap}
\end{figure}

Figure~\ref{fig:shap} shows the SHAPley values for the trained AK and SR subagents against the \code{BLineAgent} in \ref{fig:shap_bline} and \ref{fig:shap_bline_float}, and PPO against the \code{MeanderAgent} in \ref{fig:shap_meander}. Each point on these plots is a feature in a specific observation, with the colour representing the value of that feature.

All defensive agents observe the same trend in feature importance regardless of their training adversary: `adversary access' is the most important followed by `adversary scan', a trend that is also observed in Figure~\ref{fig:ablation}. Note that the PPO \code{RedMeander} defensive model doesn't use `previous action' and hence is not included in Figure~\ref{fig:shap_meander}.

We show that `adversary access' is an important part of the observation. This indicates that the defensive agents are aware that they need to remove the attackers from hosts. The importance is also seen in Figure~\ref{fig:ablation} as the most significant shifts in reward distribution occur when ablating `adversary access'. 

%
In addition `adversary scan' is of importance to the agents which is clear in~\ref{fig:ablation_bline_flaor} as the defensive agent's performance is significantly impacted in the absence of this information. This correlates with Figure~\ref{fig:shap_bline_float} as `adversary scan' has the greatest distribution of any of the SHAP values for the \code{BLineAgent} defensive agents. 
While knowledge of the `previous action' has the lowest feature importance for the agents, we argue that this is still important for these defensive agents, which, with this knowledge, outperform the baseline and PPO-only models in~\ref{fig:sub_agent_training}. For example, take the case where a defensive agent acts to remove an adversary from a host, if this action fails then the defensive agent will have to adjust its strategy.
The importance of this feature can further be seen in Figures~\ref{fig:ablation_bline} and~\ref{fig:ablation_bline_flaor}, as ablation of this feature has a non-trivial impact on the performance of the agents.
 
\section{Related Work}
\label{sec:rel-work}
The effectiveness of RL across a range of simulated and abstracted autonomous network defence scenarios is well established in the literature. Han et al.~\cite{han_reinforcement_2018} show the feasibility and resilience of RL agents under causative attacks in software defined networks. Elderman et al.~\cite{elderman_adversarial_2017} model network defence using the framework of a Markov game with incomplete information, highlighting the capabilities of even traditional RL methods (i.e., not DRL) in interactions between network attacker and defender. The hierarchical approach we build upon was first proposed by Foley et al.~\cite{foley_autonomous_2022}. Comparatively, we propose two improved controller models based on a deeper understanding of the adversary models. We also develop improved subagents providing an explainability analysis to understand what causes the agents to defend networks effectively. Other approaches to autonomous network defence include dynamic causal Bayesian optimisation~\cite{NEURIPS2021_577bcc91} as shown by Andrew et al.~\cite{dhir_causal_cyber22}. 

Several alternative network defence simulation environments have been proposed in the literature. Molina-Markham et al.~\cite{farland21} propose FARLAND which similarly to CybORG provides a hybrid simulation and emulation based environment capable, owing to a rich feature space, develops agents that can defend real-world networks. Microsoft have an experimental research platform CyberBattleSim~\cite{CyberBattleSim-github} that offers, at a high-level of abstraction, a simulation-only network defence environment based on post-breach lateral adversary movement and system exploitation. In contrast to CybORG, CyberBattleSim places greater emphasis on credential access and data collection such as simulating a GitHub project leaking credentials in the commit history. Another simulation-only environment developed by Andrew et al.~\cite{dhir_causal_cyber22} is Yawning Titan (YT). Of all the network defence environments, YT offers the greatest abstraction and omits the majority of individual host details (e.g., operating system processes, network ports) needed for emulation. 

RL has also be applied to several closely related problems. In penetration testing (i.e., exploitation which is a subset of the CybORG envionment), Yang and Liu~\cite{yangLiu-autopen22} formulate automated penetration testing in the multi-objective RL framework and demonstrate superior performance. Independently, Tran et al.~\cite{drllautope-tran21} explore hierarchical RL architectures for the same task based on their findings that decomposing large action spaces into smaller sets produces greater performing agents. In intrusion prevention, Hammar and Stadler~\cite{hammar21} demonstrate that RL is capable of intrusion prevention when formulated as a multiple stopping problem. Feng and Xu~\cite{feng_deep_2017} train a defender to protect a single device from an unknown attacker and finally, Tahsini et al.~\cite{tahsini_deepbloc_2020} use a single defender model to protect a water tank system from adversarial attacks.

\section{Conclusion}
Taking advantage of the rapidly increasing capabilities of neural networks and the advancements in RL algorithms, we present an improved approach to autonomous network defence. Beyond high performance, we place emphasis on the steps before and after training the model. 
Before training, we use a methodology to observe the adversary behaviour and inform choices in our hierarchical model. Specifically, we introduce two controller architectures, one heuristic and another bandit-based, that improve accuracy when predicting adversaries. Additionally we develop enhanced subagent architectures optimised for the specific classes of adversary.
After training, our post-hoc analysis includes a feature importance and ablation study for each specialised subagent within the complete hierarchical model. Our results shed light on each agent's decision making process and help to better understand the system as a whole. This work contributes to a less studied but equally important research direction for future works in autonomous network defence.



\section*{Acknowledgements}
The authors would like to acknowledge that research was partially funded by EPSRC grant EP/T51780X/1. 
%
%
%
 \bibliography{main}
 \newpage
\appendix
\captionsetup{width=.5\textwidth}

 \section{Hyperparameter Values}
 \label{app:hyperparam}
 Optimal, lower and upper bounds of the of the hyperparameters are shown in Table~\ref{table:hyperparam}. A uniformly sampled grid search was used to determine the optimal values.

\begin{table}[]
\centering{\footnotesize
\begin{tabular}{@{}p{3cm}|p{5cm}p{3cm}@{}}
\toprule
\multirow{2}{*}{Agent} & \multirow{2}{*}{Parameter} & \multirow{2}{*}{Value}      \\
                       &                            &                             \\ \midrule
\multirow{4}{*}{AK}    & gamma                      & 0.99                     \\
                       & network layers             & $[256, 256]$   \\    
                        & Curiosity             & Yes 
                       \\ 
                        & Beta (Curiosity)             & 0.2       
                       \\ 
                       & Eta  (Curiosity)           & 1
                       \\                      
                       & Feature Dimension  (Curiosity)           & 53
                       \\ 
                       & Learning Rate  (Curiosity)           & 0.001
                       \\ 
                       & Learning Rate             & 0.0005       
                       \\ \midrule
\multirow{4}{*}{SR}    & gamma                      & -17.710                     \\

                       & network layers             &    $[256, 256]$      \\    
                                              & Curiosity             & Yes \\ 
                        & Beta  (Curiosity)           & 0.2       
                       \\ 
                       & Eta  (Curiosity)           & 1
                       \\                      
                       & Feature Dimension   (Curiosity)          & 53
                       \\ 
                       & Learning Rate  (Curiosity)           & 0.001
                       \\ 
                         & Learning Rate & 0.0005        
                       \\ \midrule
\multirow{4}{*}{PPO}   & gamma                      & 0.99                        \\

                       & network layers             & $[256, 256, 52]$       \\    
                                              & Curiosity             & No  \\    
                                              & Learning Rate             & 0.0005    \\ \midrule
                                              
Bandits              &epsilon & 0.01\\
                                              \bottomrule
\end{tabular}}
\vspace{+2mm}
\caption{Hyperparameters.}\label{table:hyperparam}
\end{table}

\section{Extended adversary models}
\label{app:full-adv-modelz}
\captionsetup{width=.31\textwidth}
Here we provide the full action-outcome transition graphs for the \code{BLineAgent} adversary, both with and without the presence of our defensive model. Table~\ref{tab:graph_blah} provides the definitions of all the acronyms used.

\begin{table}[!h]
\centering
\begin{tabular}{@{}ll@{}}
\toprule
\multicolumn{1}{c}{Acronym} & \multicolumn{1}{c}{Definition} \\ \midrule
DRS                         & Discover Remote Systems        \\
DNS                         & Discover Network Services      \\
ERS                         & Exploit Remote Service         \\
PE                          & Privilege Escalate             \\ \bottomrule
\end{tabular}
\caption{Acronymns used in the action-outcome transition graphs.}

\label{tab:graph_blah}
\end{table}
\captionsetup{width=.5\textwidth}

\newpage
\begin{figure}
    \centering
    \includegraphics[width=1.4\textwidth, angle=90]{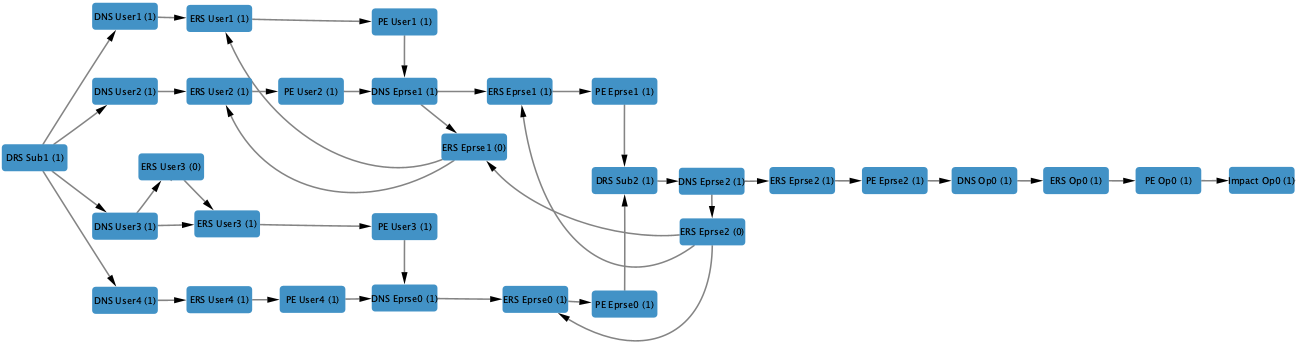}
    \caption{Action-outcome transition graph of the \code{BLineAgent} adversary without defensive action.}
    \label{fig:bline-states}
\end{figure}

\captionsetup{width=.85\textwidth}
\begin{figure}
    \centering
    \includegraphics[width=1.4\textwidth, angle=90]{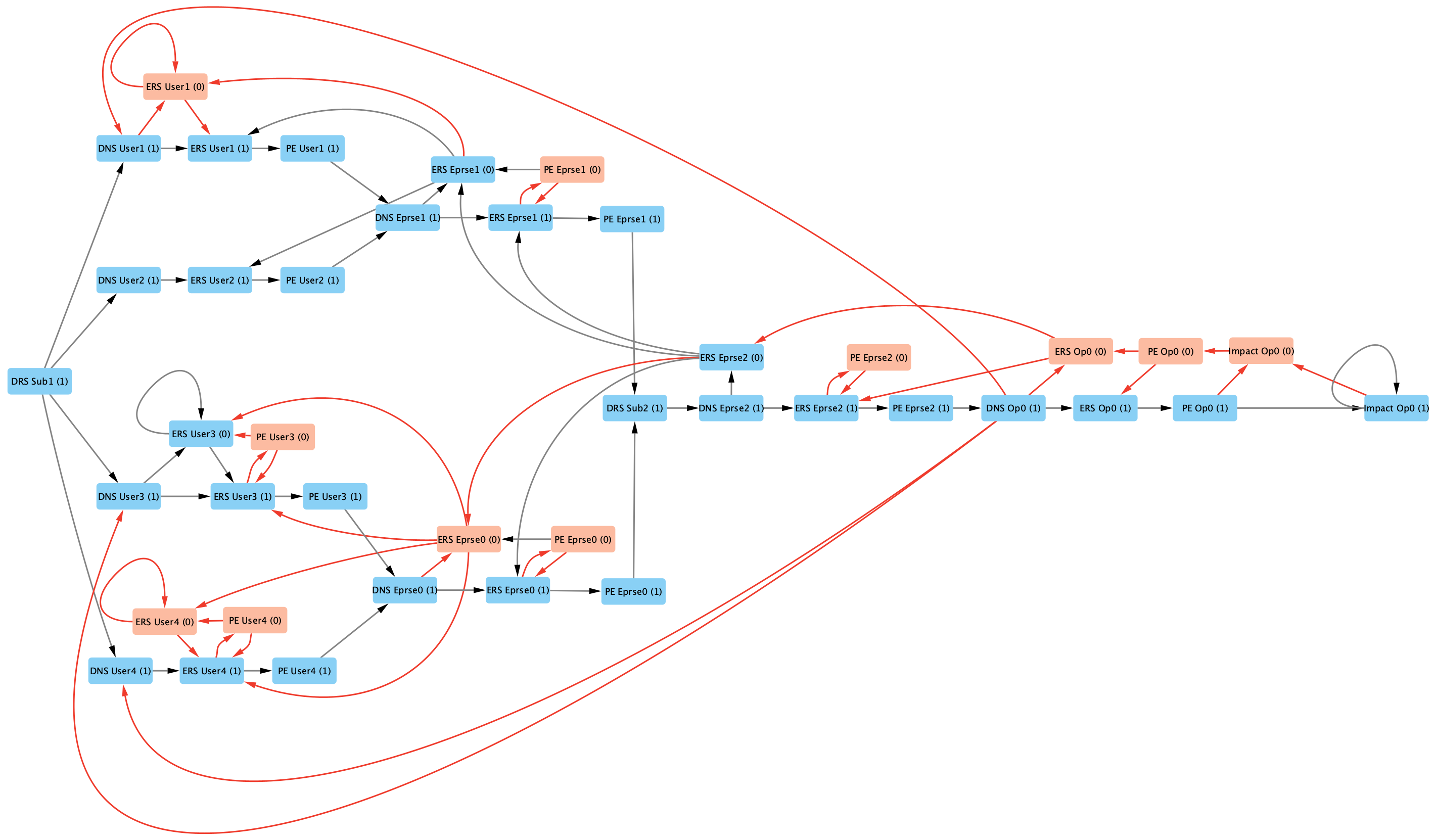}
    \caption{Action-outcome transition graph of the \code{BLineAgent} adversary faced with our fully-trained defensive model. c.f. Figure~\ref{fig:bline-states}, new states and transitions are shown in red.}
    \label{fig:bline-states-vs}
\end{figure}

\end{document}